%% file: beyondqc_stop.tex
\documentclass[pra,superscriptaddress,twocolumn,showpacs]{revtex4}
\usepackage{amssymb,amsfonts,amsmath}
\usepackage{epsfig,graphicx}\usepackage{mathrsfs}
\usepackage[hypertex,linkcolor=red]{hyperref}
\input{myQcircuit}
\newcommand{\Tr}{\operatorname{Tr}}

\newcommand{\map}[1]{\mathscr{#1}}
\newcommand{\spc}[1]{\mathsf{#1}}
\newcommand{\sH}{\spc{H}}
\newcommand{\type}[1]{\mathrm{#1}}
\newcommand{\rA}{\type{A}}
\newcommand{\rB}{\type{B}}
\newcommand{\rC}{\type{C}}
\newcommand{\rD}{\type{D}}
\newcommand{\rE}{\type{E}}

\newcommand{\rI}{\type{I}}
\newcommand{\rP}{\type{P}}
\newcommand{\rQ}{\type{Q}}

\def\Lin{{\mathsf{Lin}}}
\def\St{\mathsf{St}}
\def\QO{{\mathsf{QO}}}
\def\Chan{{\mathsf{QChan}}}
\def\Herm{{\mathsf{Herm}}}
\def\Reals{\mathbb{R}}
\newtheorem{theorem}{Theorem}
\newtheorem{definition}{Definition}
\newtheorem{lemma}{Lemma} 
\newtheorem{corollary}{Corollary} 
 
\newtheorem{proposition}{Proposition} 
\def\>{\rangle}
\def\<{\langle}
\def\map#1{\mathcal{#1}}
\begin{document}

\title{Quantum computations without definite causal structure} 

\author{Giulio Chiribella} 
\email{gchiribella@mail.tsinghua.edu.cn}
\affiliation{Institute for Interdisciplinary Information Sciences, Tsinghua University, FIT Building 1-208, Tsinghua University, Beijing, China, 100084} 
\homepage{http://iiis.tsinghua.edu.cn}
\author{Giacomo Mauro D'Ariano}
\email{dariano@unipv.it}
\affiliation{{\em QUIT Group}, Dipartimento di Fisica, Universit\`a di Pavia, and
  INFN, via Bassi 6, 27100 Pavia, Italy}
\homepage{http://www.qubit.it} 
\author{Paolo Perinotti} 
\email{paolo.perinotti@unipv.it}
\affiliation{{\em QUIT Group}, Dipartimento di Fisica, Universit\`a di Pavia, and INFN, via
  Bassi 6, 27100 Pavia, Italy} 
\homepage{http://www.qubit.it}
\author{Benoit Valiron}
\email{valiron@seas.upenn.edu}
\affiliation{CIS Department, University of Pennsylvania, 3330 Walnut St.,
  Philadelphia, PA 19104} 
\homepage{http://www.cis.upenn.edu}  
\date{\today}
\begin{abstract}
We show that quantum theory allows for transformations  of  black boxes that cannot be realized by inserting the input black boxes within a circuit in a pre-defined causal order.  The simplest example of such a transformation is the {\em classical switch of black boxes}, where two input black boxes are arranged in two different orders conditionally on the value of a classical bit.  
       The quantum version of this transformation---the {\em quantum
    switch}---produces an output circuit where the order of the connections is  controlled by a quantum bit, which becomes entangled with the circuit structure.  Simulating these transformations in a circuit with fixed causal structure requires either postselection, or an extra query to the input black boxes.       
  
\end{abstract} \maketitle

\section{Introduction}

The quantum circuit model \cite{deu89,BernsteinVazirani,yao,aharonov} is one of the most popular models of quantum computation.  In this model, information is encoded into a quantum state that evolves in time under a sequence of quantum gates. Part of the success of this model is due to its intuitive way of representing computation and to the fact that some of the best known quantum algorithms are formulated in the language of quantum circuits (see e.g. \cite{grover,simon,shor}).   

The processing of quantum states, however, is not the ultimate physical
model of computation that can be conceived within the quantum framework.  A
computation transforms an input into an output, but these do
not have to be necessarily quantum states: One can e.g.  consider a
computation where the input is a physical transformation provided as a
black box, and the output is also a transformation, obtained from the
input black box by means of suitable physical operations.   Considering these computations is quite natural from the perspective of Church's notion of computation \cite{barend}, which allows one to
compute functions of functions, rather than only functions of bits.
This type of  {\em higher-order quantum computation} is described mathematically by suitable linear maps, introduced in Refs. \cite{qca,epl} and systematically studied in Ref. \cite{comblong}. 
Clearly, higher-order quantum computation includes as a special case the processing of quantum states through time evolution.  One may wonder whether the converse holds, that is, whether every possible computation on an input black boxes can be obtained by inserting them  in a quantum circuit at definite time steps.

%Higher-order computation includes basic quantum
%information processing as a special case and is potentially more
%powerful for several information-theoretic tasks, although an assessment of its computational power in complexity-theoretic terms is still missing \cite{disclaimer}.  

In this paper we provide a counterexample, showing that there exist higher-order computations that are
admissible in principle---i.e. their existence does not lead to any
paradoxical or unphysical effect---and yet cannot be realized by inserting a single use of the input black box in a quantum circuit with fixed causal ordering of the gates.  Our counterexample consists in the execution of the program {\tt SWITCH}, where  a pair of input black boxes $\map A$ and $\map B$ are connected in two different orders ($\map B \map A$ vs. $\map A \map B$) conditionally on the value of an input bit. The impossibility of realizing the switch by simple insertion of the black boxes $\map A, \map B$ in a quantum circuit is based on the fact that  such a realization would be equivalent the realization of a time-travel machine, and therefore would violate causality.    
On the other hand, if we give up the requirement that the computation be realized by inserting  the boxes $\map A, \map B$   in a circuit \emph{in a definite order}, then there are quite simple ways to realize the switch  in a quantum laboratory, designing  quantum circuits where the geometry of the connections can be entangled with the state of a control qubit. A similar kind of macroscopic entanglement is receiving increasing
attention thanks to recent experimental breakthroughs in optomechanics \cite{pater,ted,diamonds} and in quantum optics
\cite{altepkum}.

% Having at 

The idea that computers operating without a definite causal structure
could offer advantages over conventional computers was originally
suggested by Hardy in Ref. \cite{gravitycomp}.  The first concrete
example of a task that can be accomplished only in the absence of a
pre-defined causal structure has been the execution of the program
{\tt SWITCH}, which was introduced in Ref.  \cite{v1}, of which the
present paper is an extended elaboration.  It is important to note,
however, that the program {\tt SWITCH} can be simulated by using one
extra query to the input black boxes (cf. section \ref{sec:waysaround}
of this paper). This means that quantum circuits powered by the
quantum {\tt SWITCH} are equivalent to ordinary quantum circuits in
the complexity-theoretic sense.  Nevertheless, having access to the
quantum {\tt SWITCH} offers advantages in information processing: for
example, Ref. \cite{superseq} demonstrated such an advantage in a
black box discrimination problem, while Ref. \cite{cperm} exhibited a
task where the use of the quantum {\tt SWITCH} provides a quadratic
improvement in the number of queries to the unknown black boxes.
Another concrete advantage coming from undefined causal structure came
shortly after Ref. \cite{v1}, when Oreshkov, Costa and Brukner
presented a non-local game where a causally unordered strategy offers
an advantage over causally ordered \cite{costa}. The non-causal
strategy is described by a legitimate transformation of boxes, of the
kind analyzed in this paper, but such strategy does not have a clear
operational interpretation in terms of circuits with quantum control
on the connections.
% (and, therefore, it cannot be realized using the quantum {\tt
%   SWITCH} as a subroutine).
As a consequence, it is currently unclear whether the higher-order
transformation of Ref. \cite{costa} can be also implemented by
doubling the number of queries to the input boxes.  More generally,
the physical realization of the higher-order computations described
mathematically in this paper is an important open problem for future
research. Having such a characterization is indeed the crucial step
needed to assess the computational power of the higher-order model of
quantum computation.

The paper is structured as follows:  in Section \ref{sec:qcirc} we briefly recall the framework of quantum circuits. In  Section  \ref{sec:highord}  we expose the mathematical framework of higher-order quantum transformations (a.k.a. supermaps \cite{epl,comblong}), introducing the notions of \emph{transformations on no-signalling channels} and \emph{transformations on product channels}, and providing as an example the {\tt SWITCH} transformation.    
In section \ref{sec:noswitch} we show that the {\tt SWITCH} transformation cannot be realized by inserting the input channels in a circuit, showing that such a realization would be a equivalent to the realization of a time machine.    
  In section \ref{sec:waysaround}   we discuss four ways around the no-go theorem:  having access to program states for the black boxes, using extra queries, having access to closed timelike curves, and considering probabilistic implementations of the transformation {\tt SWITCH}.   
 The possibility of re-modelling the resource of two input black boxes with control on the ordering is discussed in section    \ref{sec:remodelling}.   
Before concluding, in section  \ref{sec:quantum} we define the quantum version of the {\tt SWITCH} transformation, where the input channels $\map A$ and $\map B$ are transformed in an output quantum channel implementing a ``quantum superposition of the two circuits" $\map A \map B$ and $\map B\map A$.   
Finally,  we summarize the results of the paper in section \ref{sec:conclusions}, providing a discussion of their implications and of their relation with other works in the literature.

\section{The framework of quantum circuits}\label{sec:qcirc}  
In this section we recall a few elementary facts about the framework quantum circuits, in its version including unitary transformations as well as noisy channels (see e.g. \cite{aharonov}).   These facts will be useful to clarify in what sense higher-order transformations go beyond this model.    

In a quantum circuit quantum systems are represented by wires.  The
quantum state of the systems evolves through a sequence of quantum gates, ordered from left to right as in the following example:
\begin{equation*}
  \Qcircuit @C=.5em @R=0em @!R { 
  &\qw & \multigate{1}{\map A} & \qw & \multigate{1}{\map C}   & \qw & \qw
  \\
  &\qw & \ghost{\map A} & \multigate{1}{\map B} & \ghost{\map C} & \qw & \qw\\ 
  &\gate{f} & \qw & \ghost{\map B} & \qw & \gate{g} & \qw}
\end{equation*}
Here each wire is drawn in space, but in general the path from left to
right in the circuit does not represent a path in space: Instead, it
represents the time evolution from a computational step to the next.  In the above
example the boxes $\boxed{f}$ and $\boxed{g}$ represent transformations of single systems, e.~g. unitary gates or  noisy quantum channels.  The boxes $\map A, \map B,$ and $\map C$, instead, represent joint transformations of two systems. 

%symbol $\begin{aligned}\displaystyle\Qcircuit @C=.5em @R=0em @!R {&\qw &\ctrl{1}&\qw\\
   % &\qw&\targ&\qw}\end{aligned}$ is a C-NOT ({\em controlled-not})
%transformation: This transforms two qubits {\em jointly}, with the
%target qubit (wire with $\oplus$) undergoing the identity
%transformation if the control-qubit (wire with $\bullet$) is in the
%state $\ket{0}$ and the NOT transformation
%$\ket{0}\leftrightarrow\ket{1}$ if the control-qubit is in the state
%$\ket{1}$.  
%The symbol
%$\begin{aligned} \displaystyle\Qcircuit @C=.5em @R=0em @!R {&\qw
%    &\ctrl{1}&\qw\\ &\qw&\gate{U}&\qw}
%\end{aligned}$ is a C-U ({\em controlled-unitary}), a generalization of the control-not, with the
%transformation $U$ replacing the NOT transformation of the C-NOT.

It is worth stressing that the quantum circuit is a {\em
  computational} circuit---not a physical one: While in the physical
circuit we can have loops (e.g. when a system passes twice
through the same physical device), in the computational circuit there
are no loops (when we apply twice a transformation to the same system we
just draw two times the same box).  The computational circuit
represents the actual flow of information during the run of a
``program''. It is also important to make clear the distinction
between {\em program} and computational circuit, the former being a
set of instructions to build up the latter. In the computational circuit the ``wires'' can never
go backward, because this would mean to go {\em backward in time},
whereas in the program code we can have commands
pointing back to a previous instruction.

The framework of quantum circuits is used to evaluate the amount of computational resources used in an algorithm
({\em e.~g.} number of oracle calls, number of qubits, length of the
computation, computational space, etc.). We summarize here few basic
rules that characterize ordinary quantum circuits and the associated
resource counting.  From now on, the expression \emph{computational
  circuit} will be referred to a circuit satisfying this set of rules:
\begin{enumerate}
\item quantum systems are represented by wires;
\item a box on a single wire represents a transformation (quantum channel) on the
  corresponding system, a box on multiple wires generally describes an
  interaction between the corresponding systems;
\item input/output relations proceed from left to right and there are
  no loops in the circuit;
\item each box represents a single use of the corresponding
  transformation.
 \end{enumerate}

\section{Higher-order quantum maps}\label{sec:highord}  
In most quantum algorithms the input data are encoded in the unitary
transformation performed by a black box (the \emph{oracle}), which
represents an unknown channel, called as a subroutine during the
computation. The core of all these algorithms
describes a computation that takes as an input a certain number of
calls to the oracle, and returns as an output some classical data,
like the period of a function, or the prime factors of an integer.
From an abstract point of view, the algorithm implements a
higher-order transformation, that transforms the quantum channel
performed by the oracle into a classical output.  Generalizing this
idea, we are led to consider higher-order maps where both the input
and the output are quantum channels. These maps transform an input
oracle into a new output oracle.

The simplest example of higher-order transformations is given by the
\emph{quantum supermaps} introduced in Ref. \cite{epl}.  We now review
the main ideas in this simple case and  set up the scene for the
results of this paper.

\subsection{Notation}

In the following, we will use capital Roman letters $\rA,\rB,\dots$ to
describe types of quantum systems, such as qubits, qutrits, and so on.
Every system type $\rA$ is associated with a Hilbert space $\sH_\rA$
having dimension $d_\rA$.  The trivial system type, denoted by $\rI$,
will be associated to the trivial quantum system, with one-dimensional Hilbert space
$\sH_\rI = \mathbb C$.  The system type $\rA\rB$ will be associated to
the tensor product Hilbert space $\sH_\rA \otimes \sH_\rB$.

The linear operators from $\sH_\rA$ to $\sH_\rB$ will be
denoted by $\Lin (\sH_\rA, \sH_\rB)$ (or by $\Lin (\sH_\rA)$, if
$\sH_\rA = \sH_\rB$).
We will denote by $\St (\rA)$  the set of quantum states of system $\rA$, i.e. the set of unit trace non-negative operators in  $\Lin (\sH_\rA)$, and by  $\QO (\rA \to \rB)$ the set of {\em quantum
  operations} of type $\rA\to\rB$, i.e. the set of
trace-non-increasing completely positive (CP) maps from $\Lin
(\sH_{\rA})$ to $\Lin(\sH_{\rB})$. Similarly, we will denote by $\Chan
(\rA \to \rB)$ the set of \emph{quantum channels} of type $\rA\to\rB$,
i.e. the subset of $\QO (\rA \to \rB)$ consisting of trace-preserving
maps. Quantum operations and quantum channels of type $\rA\to\rB$ are
elements of the real vector space $\Herm (\rA \to \rB)$, consisting of
Hermitian-preserving linear maps from $\Lin (\sH_{\rA})$ to
$\Lin(\sH_{\rB})$ (see e.~g. Ref. \cite{watgut,comblong}).

\subsection{Deterministic supermaps on quantum channels}

Deterministic transformations of quantum channels where originally defined in Ref.  \cite{epl}.   A concise version of the original definition is as follows:

\begin{definition}\label{def:supermap}{\bf (Deterministic supermaps on quantum channels)}
A \emph{deterministic supermap  of type} $\Chan(\rA \to \rA')  \to \Chan(\rB \to \rB')$ is  a linear map $\map S$ from $\Herm  (\rA \to \rA')$ to $\Herm  (\rB \to \rB')$ satisfying the requirement
that  for every pair of systems $\rE, \rE'$ and for every input quantum channel $\map C \in \Chan (\rA  \rE \to \rA' \rE')$, the output  $(\map S  \otimes \map I_{\rE \to \rE'}) (\map C)$ is a quantum channel in  $\Chan (\rB \rE  \to \rB' \rE')$, where $\map I_{\rE \to \rE'}$ is the identity supermap, sending every quantum operation $\map E  \in  \QO (\rE \to \rE')$ into itself.
\end{definition}
Note in particular that for every input quantum operation $\map A \in
\QO (\rA \to \rA')$ the output $\map S (\map A)$ is a quantum
operation in $\QO (\rB \to \rB')$.

We now introduce the concepts of \emph{marginal of a channel} and
\emph{extension of a set of channels}, that besides allowing for an
intuitive re-interpretation of Def. \ref{def:supermap}, will turn out
useful when introducing supermaps on restricted sets of channels (in
Sec. \ref{subsub:rest}): the \emph{marginal on $\rA \to \rA'$} of a
given channel $\map C \in\Chan (\rA \rE \to \rA' \rE')$ relative to
state $\sigma\in\St (\rE)$ is the channel $\map C_\sigma$ defined by
\begin{align}\label{marg}
  \map C_\sigma (\rho): = \Tr_{\rE'} [\map C (\rho \otimes \sigma) ].
\end{align}

Given a set of channels $\mathsf S
\subseteq \Chan (\rA \to \rA')$ and a pair of systems $\rE, \rE'$, the
\emph{extension of $\mathsf S$ in $\Chan (\rA\rE \to \rA' \rE')$} is
the set $ \mathsf {Ext}_{\rE \to \rE'} (\mathsf S) \subseteq \Chan
(\rA\rE \to \rA' \rE')$ containing all channels $\map C $ such that
the marginal $\map C_\sigma $ in Eq. (\ref{marg}) is in $\mathsf S$
for every $\sigma \in \St (\rE)$. In formula: 
\begin{align*}
\mathsf{Ext}_{\rE \to \rE'}   ( \mathsf{S} )  :  =  \{   &  \map C \in  \Chan (\rA\rE \to \rA' \rE'  ) ~|~   ,\\
   &     \map C_\sigma \in  \mathsf{S} , \forall \sigma \in  \St(\rE)  \}.   
\end{align*} 

Using the notion of extension, Def. \ref{def:supermap} can be reformulated as follows:  

\begin{definition}{\bf (Deterministic supermaps on quantum channels: equivalent definition)}    A \emph{deterministic supermap  of type} $ \Chan (\rA \to \rA'  )  \to \Chan (\rB \to \rB')$ is  a linear map $\map S$ from $\Herm  (\rA \to \rA')$ to $\Herm  (\rB \to \rB')$ satisfying the requirement
that for every  systems $\rE, \rE'$ and for  every input quantum channel $\map C \in \mathsf {Ext}_{\rE  \to \rE'}  [\Chan(\rA \to \rA')  ] $   
   the output  $(\map S  \otimes \map I_{\rE  \to \rE'} ) (\map C)$   is a quantum channel in  $ \mathsf {Ext}_{\rE  \to \rE'} [\Chan (\rB \to \rB' )]$.  
\end{definition}

The equivalence with definition \ref{def:supermap} is obvious from the fact that the extensions  $  \mathsf {Ext}_{\rE  \to \rE'}  [\Chan(\rA \to \rA')  ] $ and in  $ \mathsf {Ext}_{\rE  \to \rE'} [\Chan (\rB \to \rB' )]$ coincide with the set of all bipartite channels $\Chan (\rA\rE \to \rA' \rE')$ and $\Chan (\rB\rE \to \rB' \rE')$, respectively.

An example of deterministic supermap is given the concatenation $\map S  (\map A)  =   \map F  (\map A  \otimes \map I_{\rC} )  \map E $, depicted as  
\begin{equation}\label{incirc}
 \begin{aligned} \Qcircuit @C=.5em @R=0em @!R { & \qw \poloFantasmaCn{\rB} & \gate {\map S(\map A)} &  \qw\poloFantasmaCn{\rB'} & \qw}
   \end{aligned}  :=  
 \begin{aligned} \Qcircuit @C=.5em @R=0em @!R { & \qw \poloFantasmaCn{\rB} & \multigate{1}{\map E} &  \qw\poloFantasmaCn{\rA} & \gate{\map A}  & \qw \poloFantasmaCn{\rA'}  &  \multigate{1}{\map F}  &  \qw  \poloFantasmaCn{\rB'} & \qw\\
    & & \pureghost{\map E} &  \qw\poloFantasmaCn{\rC} & \qw& \qw &  \ghost {\map F}  &  &}
   \end{aligned}
\end{equation}
where $\rC$ is a suitable quantum system, and $\map E  \in \Chan(\rB  \to \rA  \rC)$  and $\map F  \in  \Chan (\rA'\rC  \to \rB')$ are suitable quantum channels.   By definition, the transformations of the form of Eq. (\ref{incirc}) are exactly those that can be obtained by inserting a single use of the input channel $\map A$ inside a quantum circuit.   One of the results of Ref.  \cite{epl} is that every linear map satisfying the requirements of Def. \ref{def:supermap} is a concatenation of the above form: deterministic supermaps on arbitrary channels can always be realized by insertion in a suitable quantum circuit. 
This means that if we want to find a counterexample of higher-order transformation that cannot be realized by insertion in a quantum circuit we have to search in a different family of supermaps.

\subsection{Generalizations: hierarchy of higher-order maps and supermaps on restricted sets of channels} \label{subsub:rest}

The example of supermaps on quantum channels is the key for two important generalizations:  
\begin{enumerate}
\item \emph{Hierarchy of higher-order maps:}   lifting Def. \ref{def:supermap} to the next level, we can define linear maps that transform quantum supermaps into quantum supermaps, preserving normalization when acting locally on one side of a bipartite input.  Iterating this procedure,  we then obtain an infinite hierarchy of higher-order quantum maps. 
\item \emph{Supermaps that transform restricted sets of quantum channels:}  instead of imposing that every channel is sent to a channel as in Def. \ref{def:supermap}, we can define supermaps that transform a restricted set  of quantum channels (e.g. the no-signalling ones)  to another, sending elements in the extension of the former into elements in the extension of the latter. 
\end{enumerate}

The complete characterization and the physical interpretation of these
new quantum maps is a difficult open problem.  Regarding the
generalization 1, part of the hierarchy of higher-order maps has been
characterized in Ref. \cite{comblong}. Precisely, Ref. \cite{comblong}
characterized the types of higher-order maps that can be realized
within the quantum circuit framework.   

Regarding the generalization 2, a more formal definition of supermaps acting on a restricted set of channels can be given as follows:

\begin{definition}\label{def:general}{\bf (Deterministic supermaps on a restricted set of quantum channels)}  Let $  \mathsf S_A  \subseteq \Chan (\rA \to \rA')$ and $\mathsf S_B\subseteq \Chan (\rB \to \rB')$  be two subsets of quantum channels. A \emph{deterministic supermap  of type} $ \mathsf S_{A}  \to \mathsf S_B $    is  a linear map $\map S$ from $\Herm  (\rA \to \rA')$ to $\Herm  (\rB \to \rB')$ satisfying the requirement
that for every  systems $\rE, \rE'$ and for  every input quantum channel $\map C \in \mathsf {Ext}_{\rE  \to \rE'}  [ \mathsf S_A  ] $   
   the output  $(\map S  \otimes \map I_{\rE  \to \rE'} ) (\map C)$   is a quantum channel in  $ \mathsf {Ext}_{\rE  \to \rE'} [ \mathsf S_B]$.  
\end{definition}

Several results that are useful for the characterization of supermaps
on restricted sets of channels have been recently found by Jen\^cov\'a
\cite{genchan}.   However, also in this case the physical realizability of these supermaps is an open problem.   
In this paper we will focus on supermaps on \emph{no-signalling channels}, which is one of the most interesting classes of supermaps on restricted sets of channels.

%However, the application of these results to establish  the realization of supermaps transforming no-signalling channels is far from straightforward.

 \subsection{Choi representation of higher-order maps}    
The simplest way to study higher-order maps is via the Choi isomorphism, namely the  one-to-one
correspondence between quantum operations $\map Q  \in  \QO(\rA \to \rB)$ and positive
operators $Q  \in  \Lin (\sH_{\rB}  \otimes \sH_\rA)$ given by the relations  
\begin{align}
  Q& =  (\map Q\otimes \map I_\rA )(|  I_\rA\>\<I_\rA | ),\nonumber\\
  \map Q(\rho)  &=  \Tr_\rA[(I_\rB\otimes\rho^T )Q]  \qquad \forall \rho \in \Lin(\sH_\rA) , 
\end{align}
where $\map I_\rA$ denotes the identity map on $\Lin(\sH_\rA)$,
$\sH_\rA^{\otimes2}\ni|I_\rA \>:=\sum_{n=1}^{d_\rA}|n\>\otimes|n\>$,
$\Tr_\rA$ denotes the partial trace on $\sH_\rA$, and $\rho^T$ denotes
the transpose of $\rho$ in the basis $\{|n\>\}_{n=1}^{d_\rA}$ used in
the definition of $|I\>$.

Via the Choi isomorphism, we have that a linear map  $\map S:  \Herm(\rA \to \rA')  \to \Herm(\rB \to \rB')$ can be equivalently represented by a linear map  $\widetilde {\map S}$ from $\Lin (\sH_{\rA'}  \otimes \sH_\rA)$  to  $\Lin (\sH_{\rB'}  \otimes \sH_\rB)$, uniquely defined by the relation \cite{epl}
\begin{align}\label{choio}
\map B =  \map S (\map A)    \Longleftrightarrow  B  =  \widetilde {\map S}  (A)   \qquad   & \forall \map A \in \QO (\rA \to \rA') \\
\nonumber &\forall \map B  \in \QO (\rB \to \rB').  
 \end{align} 
  
Now, the supermaps introduced in Def. \ref{def:general} are not arbitrary linear maps: they send quantum channels to quantum channels also when acting locally on suitable bipartite extensions.  This property of a supermap $\map S$  forces the complete positivity of the map $\widetilde {\map S}$   in the Choi representation.   This fact is easy to show when the set of input channels for $\map S$ contains an \emph{internal channel}  $\map C_0 $:  
  
\begin{definition} 
 A channel $\map C_0 \in  \Chan (\rA \to \rA') $ is  \emph{internal} if for every quantum operation $\map Q  \in  \QO(\rA \to \rA')$  there exists a scaling factor $\lambda >  0$  such that the map $ \map C_0  -\lambda  \map Q $ is completely positive.
 \end{definition}
     
The completely depolarizing channel, defined by  $\map C_0 (\rho) : =  \Tr[\rho]  \frac I{d_{\rA'}} $ is an example of internal channel.             

With this definition, we are ready to state  the property of complete positivity for supermaps: 
\begin{theorem}[Complete positivity of supermaps]\label{theo:cp}
Let $\mathsf S_A \subseteq  \Chan (\rA \to \rA')$ and  $\mathsf S_B \subseteq  \Herm (\rB \to \rB')$ be two restricted sets of quantum channels, with the property that $\mathsf S_A$ contains an internal channel $\map C_0$.  
Let $\map S:   \Herm (\rA \to \rA') \to  \Herm (\rB \to \rB')$ be a supermap of type  $\mathsf S_A  \to \mathsf S_B$.
Then, in the Choi representation, the map $\widetilde {\map S}$ is completely positive.   
\end{theorem}

The proof of the theorem is given in appendix \ref{app:one}.   

As an immediate implication,  theorem \ref{theo:cp} implies that supermaps on arbitrary quantum channels are represented by completely positive maps in the Choi picture (simply because the set of all quantum channels includes the completely depolarizing channel). Similarly, all the types of supermaps considered in this paper will satisfy the hypothesis of theorem \ref{theo:cp} and hence will be described by completely positive maps $\widetilde{\map S}$ in the Choi picture.  

Like every completely positive map, a supermap $\widetilde{\map S}$ can be written in the Kraus form $\widetilde {\map S} (A)  =  \sum_n   S_n  A S_n^\dag$.  
Complete positivity is a very powerful property, which in certain situations allows one to define a supermap uniquely  by only specifying its action only on quantum channels. 

\subsection{Deterministic supermaps on no-signalling channels}

In the rest of the paper we will focus on supermaps that transform a restricted set of quantum channels, namely the set of (bipartite) \emph{no-signalling channels}.    We recall that a bipartite channel in  $\Chan (\rA \rB  \to  \rA' \rB')$ is no-signalling if there exist two channels $\map A  \in  \Chan (\rA \to \rA')$ and $\map B \in  \Chan (\rB \to \rB')$ such that
\begin{align*}
\Tr_{\rA'}  [\map C(\rho)]   &  =  \map B  (\Tr_{\rA}  [\rho] )  \qquad \forall  \rho  \in \Lin (\sH_\rA  \otimes \sH_\rB)   \\
\Tr_{\rB'}  [\map C(\rho)]   &  =  \map  A  (\Tr_{\rB}  [\rho] )  \qquad \forall  \rho  \in \Lin (\sH_\rA  \otimes \sH_\rB)    
\end{align*} 
(see e.g. \cite{pianihoro}).

Following the general definition \ref{def:general}, we can define supermaps on no-signalling channels as follows: 

\begin{definition}\label{def:supermapns}
Let $\mathsf {NS}(\rA\rB \to \rA'\rB')$  denote the set of no-signalling channels in $\Chan (\rA \rB \to \rA' \rB')$.     A \emph{deterministic supermap  of type $ {\mathsf{NS}} (\rA  \rB\to \rA'  \rB')  \to \Chan(\rC \to \rC')$} is  a linear map $\map S$ from $\Herm  (\rA\rB \to \rA'\rB')$ to $\Herm  (\rC \to \rC')$ satisfying the requirement
that for every  systems $\rE, \rE'$ and for  every input quantum channel $\map C \in \mathsf {Ext}_{\rE  \to \rE'}  [\mathsf {NS}(\rA\rB \to \rA'\rB')  ] $   
   the output  $(\map S  \otimes \map I_{\rE  \to \rE'} ) (\map C)$   is a quantum channel in  $ \mathsf {Ext}_{\rE  \to \rE'} [ \Chan(\rC \to \rC')]\equiv\Chan (\rC\rE \to \rC' \rE')$.  
\end{definition}

Note that the normalization condition in Def. \ref{def:supermapns} is weaker than the one in Def. \ref{def:supermap}, because the latter requires the output to be a channel whenever the input is a channel, while the former requires the output to be a channel \emph{only if the input channel is no-signalling.} 
As a consequence, the set of supermaps on no-signalling channels is larger  than the set of ordinary supermaps described by Def. \ref{def:supermap}.  
Moreover, since  the ordinary supermaps are all and only those transformations that can be implemented by inserting the input channel in a suitable circuit \cite{epl}, all the supermaps on no-signalling channels  which are outside the set of ordinary supermaps cannot be implemented in the circuit model (that is, cannot be implemented by inserting one use of the input channel inside a quantum circuit).   An example of this kind is the switch supermap, introduced in Ref. \cite{v1}  and discussed extensively  in the next section of this paper.  
Another example of supermap that cannot be realized by insertion in a quantum circuit  is given by the map defined by Oreshkov, Costa, and Brukner \cite{costa},  whose input is the set of  no-signalling channels in $\Chan (\rA \rB  \to \rA'\rB')$,  $\sH_\rA  \simeq \sH_\rB \simeq \sH_{\rA'}  \simeq \sH_{\rB'}  \simeq \mathbb C^2$.    

In the Choi picture, a supermap $\map S$ on no-signalling channels is described by a completely positive map $\widetilde {\map S}$.   Complete positivity can be easily proved from theorem \ref{theo:cp}, using the fact that the depolarizing channel   is a no-signalling channel.

 \subsection{Alternative characterization of supermaps on no-signalling channels}

Supermaps on no-signalling channels  can be equivalently characterized as \emph{supermaps on product channels}, according to the following definition:

\begin{definition}\label{def:supermapprod}{\bf (Supermaps on product channels)}
Let ${\mathsf {PROD}}(\rA\rB \to \rA'\rB')  =  \{  \map A \otimes \map B ~,~  \map A \in\Chan (\rA \to \rA'), \map B \in \Chan(\rB \to \rB'\}$  denote the set of product channels in $\Chan (\rA \rB \to \rA' \rB')$.     A \emph{deterministic supermap on product channels of type} $ {\mathsf{PROD}} (\rA  \rB\to \rA'  \rB')  \to \Chan(\rC \to \rC')$ is  a linear map $\map S$ from $\Herm  (\rA\rB \to \rA'\rB')$ to $\Herm  (\rC \to \rC')$ satisfying the requirement
that for every  systems $\rE, \rE'$ and for  every input quantum channel in the extension set $\map C \in \mathsf {Ext}_{\rE  \to \rE'}  [{\mathsf {PROD}}(\rA\rB \to \rA'\rB')  ] $   
   the output  $(\map S  \otimes \map I_{\rE  \to \rE'} ) (\map C)$   is a quantum channel in  $ \mathsf {Ext}_{\rE  \to \rE'}  [ \Chan(\rC \to \rC')]  \equiv \Chan (\rC\rE \to \rC' \rE')$.  
\end{definition}

Obviously, product channels are a special case of no-signalling channels. Hence, every supermap on no-signalling channels is also a supermap on product channels.  Less trivially, we will now show that also the converse is true:  the set of supermaps on no-signalling channels coincides with the set of supermaps on product channels.  
This result is useful because it is much easier to check that a supermap satisfies the definition on product channels, instead of the one on general no-signalling channels.   

\begin{theorem}\label{theo:alternative}{\bf (Supermaps on no-signalling channels $=$ supermaps on product channels)}
The set of deterministic supermaps of type   $ {\mathsf{NS}} (\rA  \rB\to \rA'  \rB')  \to \Chan(\rC \to \rC')$ coincides with the set of deterministic supermaps of type $ {\mathsf{PROD}} (\rA  \rB\to \rA'  \rB')  \to \Chan(\rC \to \rC')$.  
Moreover, the correspondence between elements of the two sets is one-to one: if two supermaps act in the same way on product channels, then they act in the same way on arbitrary no-signalling channels.
\end{theorem}

In oder to prove the theorem we need to collect  a few ingredients.   The first ingredient is an alternative characterization of the set of no-signalling channels as affine combinations of product channels.   Such a characterization can be easily obtained building on a result of Ref.\cite{gut}: 
  
\begin{lemma}\label{lem:nosig}{\bf (No-signalling channels are affine combinations of product channels)}   
A quantum channel  $\map C  \in  \Chan (\rA \rB  \to \rA'  \rB'  )$ is no-signalling if and only if it is an affine combination of the form   $\map C  =   \sum_i   \lambda_i     ~  \map F_i  \otimes \map G_i$, with  $\lambda_i \in \Reals$, $\map F_i \in  \Chan (\rA \to \rA') $,  $\map G_i \in  \Chan (\rB \to \rB') $ for every $i$ and $\sum_i \lambda_i = 1 $. 
\end{lemma}  

{\bf Proof.}  
 Ref. \cite{gut}  proved that $\map C$ is a no-signalling channel if and only if $\map C  =   \sum_i   \lambda_i     ~  \map F_i  \otimes \map G_i$, where   $\map F_i   \in  \Herm (\rA \to \rA') $,    $\map G_i \in  \Herm (\rB \to \rB') $ are trace-preserving maps and $\lambda_i \in \mathbb R$   for every $i$. Clearly, the trace-preserving property of $\map C, \map F_i$ and $\map G_i$ forces the linear combination to be affine, namely $\sum_i \lambda_i = 1$.    
 Now, to prove our thesis we only need to observe that every Hermitian-preserving trace-preserving map is an affine combination of quantum channels.  The proof of this fact is proven in the following lemma  \ref{lem:tracepresmap}.   $\blacksquare$ 
 
 \begin{lemma}\label{lem:tracepresmap}{\bf (Hermitian-preserving trace-preserving maps are  affine combinations of quantum channels)}   
Every Hermitian-preserving trace-preserving map  $\map L  \in  \Herm (\rA   \to \rA'    )$ can  be written in the form      $\map L  =   \theta   \map C_+  +   (1  -\theta)   \map C_-$, where  $\map C_\pm \in  \Chan (\rA   \to \rA'    ) $ are quantum channels and  $\theta  \ge 0$.  
\end{lemma}    

   {\bf Proof.} Consider an arbitrary Hermitian-preserving and trace-preserving linear map $\map L \in  \Herm(\rC  \to \rC')$.  Write it as $\map L  =     \map L_+  -     \map L_-$, where $\map L_{\pm} $ are completely positive maps in $ \Herm (\rC \to \rC')$.  Since $\map L$ is trace-preserving, we have  
 \begin{align}\label{ffff}
\Tr[\rho] = \Tr[ \map L_+  (\rho)]  - \Tr[\map L_-(\rho)] \qquad \forall \rho\in\St(\rC).
\end{align} 
 By defining $\theta :  = \max_{\rho  \in  \St (\rC)}   \Tr[\map L_+(\rho)]   
$ we can now introduce the maps $\map C_+$ and $\map C_-$ via the relation
\begin{align*}
\theta \map C_+  (\rho) &:  =  \map L_+ (\rho)  +  \frac {I_{\rC'}}{d_{\rC'}} ~ (  \theta \Tr [\rho]  -  \Tr[\map L_+(\rho)])   \\
(\theta-1)  \map C_-  (\rho)  &  :  =   \map L_- (\rho)  +  \frac {I_{\rC'}}{d_{\rC'}} ~ (  \theta \Tr [\rho]  -  \Tr[\map L_+(\rho)]) ,  \end{align*}  
for every state $\rho\in\St (\rC)$.  Using Eq. (\ref{ffff}) and the definition of $\theta$ it is immediate to check that $\map C_\pm$ are completely positive and trace-preserving, that is, they are quantum channels.   Moreover, by construction $\map L $ can be expressed as  a linear combination
 $\map L  =   \theta   \map C_+  +   (1  -\theta)   \map C_-$, thus proving the thesis. 
 $\blacksquare$  
   
\medskip 
 
 Lemma \ref{lem:nosig} implies the following corollary: 
 
%can be characterized as completely positive bilinear maps sending pairs of quantum channels into quantum channels:  
\begin{corollary}\label{cor:onproducts}{\bf (The action of a linear map on no-signalling channels is completely identified by its action on product channels)}  
%The set of supermaps of type  $ {\mathsf{NS}} (\rA  \rB\to \rA'  \rB')  \to (\rC \to \rC')$   is completely identified by its action on product channels $  \map C  =\map A  \otimes \map B ,  \map A \in  \Chan (\rA \to \rA'), \map B \in  \Chan (\rB \to \rB')$.  %with the set of  bilinear maps $\map S$ from $\Herm  (\rA\rB \to \rA'\rB')$ to $\Herm  (\rC \to \rC')$
%In other words, f
Let $\map S, \map S' $ be two linear maps from $\Herm  (\rA\rB \to \rA'\rB')$ to $\Herm  (\rC \to \rC')$. Then, the  following condition holds 
\begin{align*}
\map S (  \map A  \otimes \map B )  &=   \map S' (  \map A  \otimes \map B )  ,  \qquad \begin{array}{l}  \forall  \map A \in  \Chan (\rA \to \rA') \\
  \forall \map B \in  \Chan (\rB \to \rB') \end{array}  \\
  &\\
  &   \Longrightarrow  \map S (\map C) = \map S' (\map C)  \qquad \forall \map C \in  {\mathsf{NS}} (\rA  \rB\to \rA'  \rB')    &  
\end{align*}
\end{corollary}

Now, to prove theorem \ref{theo:alternative} it remains to take care of complete positivity: we have to ensure that the output of a supermap on product channels is completely positive even when the supermap is applied to a no-signalling channel.  
In fact, thanks to theorem \ref{theo:cp}, we are in position to prove  a much stronger result:  supermaps on quantum channels produce a completely positive output \emph{even when the input is an arbitrary completely positive map:}  

\begin{lemma}\label{lem:completeposprod}{\bf (Supermaps on product channels are completely positive)} 
 Let    $\map S $ be a supermap of type $\mathsf{Prod} (\rA\rB \to \rA'\rB') \to \Chan (\rC \to \rC')$.  Then,  for every pair of systems $\rE, \rE'$  and for for every quantum operation $\map Q \in  \QO (\rA \rB \rE \to\rA' \rB' \rE')$ the map $(\map S\otimes \map I_{\rE \to \rE'})  (\map Q) $ is completely positive. 
\end{lemma} 

{\bf Proof.}  The set of product channels contains the internal channel $\map C_0  =  \map C_{0,A}  \otimes \map C_{0,B}$, where $\map C_{0,A}  (\rho)  =  \Tr[\rho]  I_{\rA'}/d_{\rA'}$ and $\map C_{0,B}  (\rho)  =  \Tr[\rho]  I_{\rB'}/d_{\rB'}$ are depolarizing channels.   Hence, thanks to theorem \ref{theo:cp}, the map $\widetilde{\map S}$ is completely positive.    
Translating back from the Choi picture, this means that $(\map S  \otimes  \map I_{\rE  \to \rE'})$ sends completely positive maps to completely positive maps.  $\blacksquare$  

\medskip 

We can finally conclude with the proof of Theorem \ref{theo:alternative}:  
\noindent {\bf Proof of theorem \ref{theo:alternative}.}   Since supermaps on no-signalling channels are automatically supermaps on product channels, to prove that the two sets are the same we only need to prove the converse inclusion: we need to prove that supermaps on product channels are necessarily supermaps on no-signalling channels.  
Let $\map S$ be a supermap on product channels and let $\map C   \in  \mathsf {Ext}  [  \mathsf{NS} ]  (\rA \rB  \to \rA' \rB')$  the extension of some no-signalling (not necessarily product) channel. 
Then, by lemma \ref{lem:completeposprod} the map $\map C'  := (\map S\otimes \map I_{\rE \to \rE'})  (\map C) $ is completely positive.    
We now have to guarantee that $\map C'$ is trace-preserving.   
 To this purpose, note that  for every pair of  quantum states $\rho\in\St (\rA \rB) ,  \sigma  \in \St (\rE)$ we have 
 \begin{align*}
 \Tr[   \map C'  (  \rho  \otimes \sigma)]  &  =   \Tr\{ [   \map S   (   \map C_\sigma ) ] (\rho)     \},
 \end{align*}
 where we  $\map C_\sigma$  is the channel defined by $\map C_{\sigma}  (\rho)  :  =  \map C  (  \rho  \otimes \sigma )$.   
Since $\map C$ is the extension of a no-signalling channel, the channel $\map C_\sigma$ is no-signalling.  Then, by lemma \ref{lem:nosig}, we can write $\map C_{\sigma}$ as an affine combination of product channels  $\map C_\sigma  =  \sum_{i}  \lambda_{i,\sigma}   ~ (\map A_{i,\sigma} \otimes \map B_{i,\sigma})$.    Now, since $\map S$ is a supermap on product channels,  $\map S  ( \map A_{i,\sigma}  \otimes \map B_{i,\sigma})$ is a channel for every $i$, and, in particular, it is trace-preserving.  We then conclude      
 \begin{align*}
 \Tr[   \map C'  (  \rho  \otimes \sigma)]  &  =  \sum_i \lambda_{i,\sigma}  ~   \Tr\{ [   \map S   (   \map A_{i,\sigma}  \otimes \map B_{i ,\sigma} ) ] (\rho)     \}  \\
 &  =     \sum_i   \lambda_{i,\sigma}  =  1.
 \end{align*}
Since product states are a spanning set, the above equation proves that $\map C'  =  (\map S\otimes \map I_{\rE \to \rE'})  (\map C) $ is a trace-preserving.   Hence, we have proved that $\map S$ is a supermap on no-signalling channels.  Finally, the correspondence between supermaps on product channels and supermaps on no-signalling channels is 1-to-1:   if two supermaps $\map S , \map S'$  on no-signalling channels satisfy $\map S  (\map A  \otimes \map B)  = \map S'  (\map A  \otimes \map B)   $ for arbitrary product channels, then $\map S  =  \map S'$.     $\blacksquare$

\subsection{The switch supermap}\label{subsect:switch}

Here we  show an example of supermap on no-signalling channels that cannot
be realized by inserting the input  in a given quantum
circuit.  The example is given by the \emph{switch supermap} $\map Z$, which is defined as a supermap of type ${\mathsf{NS  }}  ( \rA \rB \to \rA'\rB ')  \to  \Chan (\rC \to \rC')$ with $\rA = \rB  = \rA'=\rB'   = \rC' = \mathbb C^2$ and  $\rC  = \rA\rQ  $, where $\rQ  =  \mathbb C^2$.   
The supermap $\map Z$ transforms an arbitrary pair of quantum channels $\map A \in \Chan(\rA\to
\rA'),\map B \in  \Chan(\rB\to
\rB')$ into the classically-controlled channel that performs either the
transformation $\map B \map A$ or the transformation $\map A \map B$
conditionally on the outcome of a measurement on the control qubit $\rQ$.  
Precisely, the output of the supermap is the channel $\map Z (\map
A\otimes \map B) \in \Chan (\rA\rQ \to \rA)$   defined by
\begin{align}\label{zeta}
\map Z (\map A\otimes  \map B)  (\rho) : =   \map B \map A  (   \< 0    |_\rQ  \rho  |0\>_\rQ  )   +   \map A \map B  (   \< 1    |_\rQ  \rho  |1\>_\rQ  ) , 
\end{align}
where $\<i|_\rQ \rho|i\>_\rQ$ is the state of system $\rA$ conditional to the
outcome $i$ of an orthogonal measurement on the control qubit $\rQ$.

Equation (\ref{zeta}) defines the action of the linear map $\map Z$ on the set of product channels, and, by linearity, also on the set of no-signalling channels (cf. lemma \ref{lem:nosig}).  
If $\map Z$ where just a linear map, then we would be free to choose how to define it outside the subspace spanned by no-signalling channels.  However, since we require $\map Z$ to be a \emph{supermap on no-signalling channels}, $\map Z$ has to satisfy the additional constraint of complete positivity.  
Surprisingly, it is possible to show that Eq. (\ref{zeta}) combined with complete positivity determines the action of $\map Z$ \emph{on arbitrary quantum operations}.  
  
\begin{lemma}\label{lem:switchext}
The switch supermap  $\map Z   $ is uniquely defined by  Eq. (\ref{zeta}).  In particular, for two arbitrary quantum operations  $\map Q_A  \in\QO (\rA  \to \rA')$ and $\map Q_B \in \QO (\rB  \to \rB')$ one has 
\begin{align*}
\map Z (\map Q_A\otimes  \map Q_B)  (\rho)  =   \map Q_B \map Q_A  (   \< 0    |_\rQ  \rho  |0\>_\rQ  )   +   \map Q_A \map Q_B  (   \< 1    |_\rQ  \rho  |1\>_\rQ  ) .
\end{align*}  
\end{lemma}

{\bf Proof.}  Eq. (\ref{zeta}) is equivalent to 
\begin{align}\label{suiccio}
\map Z  ( \map A \otimes \map B)  =     \map P_0  \otimes \map Z^{(0)}  & ( \map A \otimes \map   B )     +     \map P_1  \otimes \map Z^{(1)}   ( A \otimes B)   , 
%\nonumber  & \forall \map A  \in  \Chan (\rA \to \rA'),  \forall \map B \in  \Chan (\rB \to \rB')    
\end{align}  
where   $\map P_i  (\rho)  =  \<i|_\rQ  \rho |i\>_\rQ$, $i=0,1$ are the quantum operations representing the measurement on the control qubit $\rC$, and $\map Z^{(i)}  :  \Herm (\rA \rB \to \rA' \rB')  \to \Herm (\rA  \to \rA)$, $i  =  0,1$ are two linear maps such that   
\begin{align}
\label{zero}  \map Z^{(0)} ( \map A\otimes \map B) & =  \map B \map A   \\
\label{uno}
 \map Z^{(1)} ( \map A \otimes \map B)  &=  \map A \map B ,
\end{align}   
for every pair of quantum channels $\map A  \in  \Chan (\rA \to \rA')$  and   $\map B \in  \Chan (\rB \to \rB')$.  

Clearly, $\map Z$ is a supermap on no-signalling channels if and only if $\map Z^{(0)}$ and $\map Z^{(1)}$ are both supermaps on no-signalling channels.   We now show that,  due to complete positivity, Eqs. (\ref{zero})  and (\ref{uno}) are sufficient to identify the supermaps $\map Z^{(0)}$ and  $\map Z^{(1)}$ uniquely.  To this purpose, we use the Choi representation of Eq. (\ref{choio}), where each $\map Z^{(i)}$  $i=  0,1$  is represented by a completely positive linear map  $\widetilde {\map Z}^{(i)}:   \Lin  (  \sH_{\rA'}  \otimes \sH_{\rA}  \otimes \sH_{\rB'} \otimes \sH_{\rB}) \to \Lin  (\sH_{\rA}  \otimes \sH_\rA )$.  

We now show that Eq. (\ref{zero})  completely determines the map $\widetilde {\map  Z}^{(0)}$ (and hence $\map Z^{(0)}$, since the correspondence $ \map Z^{(0)}  \leftrightarrow \widetilde{\map Z}^{(0)}$ is one-to-one).    
Let us consider the case when $\map A$
and $\map B$ are both unitary channels. For a unitary channel $\map U(\rho)=U\rho
U^\dag$, the Choi operator is the rank-one operator $|U\rangle\langle U|$, where $|U\>$ is the vector defined by $|U\>  : =  (U  \otimes I )  |I\>$.    Using  Eq. (\ref{zero}) we then obtain  
\begin{align*}
\map Z^{(0)}  ( |U\>\<U|  \otimes |V\>\<  V| )    =  |UV  \>\<  UV| ,
 \end{align*} 
for every unitary operators $U$ and $V$.   Writing the map $\widetilde{\map Z}^{(0)}$  in the Kraus form $\widetilde{\map Z}^{(0)} (C)  =  \sum_n   Z_n^{(0)}   C  Z_n^{(0)\dag} $  (recall that  $\widetilde{ \map Z_0}$ is completely positive by theorem  \ref{theo:cp}]  ),   we then get
\begin{align}\label{vu}
\sum_n   Z_n^{(0)}    (  |U\>\<U|  \otimes |V\>\<  V|)    Z_n^{(0)\dag}   =  |UV  \>\<  UV| ,    
\end{align}
for every unitary operators $U$ and $V$.    Hence, for every $n$ we must have
\begin{align}\label{eq}
  Z^{(0)}_n |U\rangle|V\rangle  =\alpha^{(0)}_{n, U, V} |UV\rangle \end{align}
 for some complex number $\alpha^{(0)}_{n, U, V}$, which possibly depends on $U$ and $V$.   Note that  Eq. (\ref{vu})  imposes $\sum_n   \left | \alpha^{(0)}_{n, U, V} \right|^2   =  1$ for every unitaries $U, V$. 
 
 Applying Eq. (\ref{vu})  in the case where $U$ and $V$ are Pauli matrices  $\{  \sigma_\mu\}_{\mu =  0}^3$, $\sigma_0  =  I$,  $\{\sigma_1 , \sigma_2, \sigma_3\} \equiv  \{\sigma_x, \sigma_y, \sigma_z \}$, we have 
\begin{align}\label{munu}
 Z^{(0)}_n |\sigma_\mu  \rangle|\sigma_\nu  \rangle=\alpha^{(0)}_{n, \mu, \nu}|\sigma_\mu  \sigma_\nu \rangle \end{align}   
 
Now we show that $\alpha^{(0)}_{n ,\mu,\nu} $ is independent of $\mu$
and $\nu$, say $\alpha_{n, U, V}  \equiv  \alpha_n, \forall \mu,\nu  \in  \{  0, 1,2,3\}$.  To see that that $\alpha^{(0)}_{n ,\mu,\nu} $ is independent of $\mu$
and $\nu$, consider the unitary $U = \frac 12 \sum_\mu \omega_\mu ~
\sigma_\mu $, where $ \omega_ 0 = 1 $ and $\omega_\mu = i$ for $\mu =
1,2,3$.  Eq. (\ref{eq}) then gives
\begin{align*}
  Z^{(0)}_n |\sigma_\mu\rangle|U\rangle   & =\alpha^{(0)}_{n, \mu, U} |\sigma_\mu U\rangle\\
  &= \sum_{\nu} \frac{\alpha^{(0)}_{n, \mu, U} ~\omega_\nu }2
  |\sigma_\mu \sigma_\nu \rangle ,
\end{align*} 
whereas linearity and Eq. (\ref{munu}) give
\begin{align*}
  Z^{(0)}_n |\sigma_\mu\rangle|U\rangle= \sum_\nu
  \frac{\alpha^{(0)}_{n, \mu, \nu} ~\omega_\nu} 2 |\sigma_\mu
  \sigma_\nu \rangle .
\end{align*}   
Hence, by comparison we obtain $ \alpha^{(0)}_{n, \mu, \nu} =
\alpha^{(0)}_{n, \mu, U} $ for every $\mu, \nu$.  This shows that
$\alpha^{(0)}_{n, \mu, \nu} $ cannot depend on $\nu$.  Repeating the
same argument for $Z^{(0)}_n(|U\rangle|\sigma_\nu\>)$, we can also
prove that $\alpha^{(0)}_{n, \mu, \nu} $ cannot depend on $\mu$.  In
conclusion, we have $\alpha^{(0)}_{n, \mu, \nu} = \alpha^{(0)}_{n}$
for every $n,\mu,\nu$.  

Using linearity and the completeness of the Pauli matrices
$\{ \sigma_\mu\}_{\mu = 0}^3$ in the space of linear operators this
implies that 
\begin{align*}
Z^{(0)}_n  |A\>  |B\>  =   \alpha_n  |AB\>  \qquad \forall A , B   \in  \Lin (\mathbb{C}^2)
\end{align*}
and, therefore $\widetilde{\map Z}^{(0)} ( |A\>\<A| \otimes |B\>\<B|)
= |A B\>\< AB|$ for every $A, B \in \Lin (\mathbb C^2)$.
Finally, using the normalization condition $\sum_n
\left|\alpha^{(0)}_{n}\right|^2 = 1$, we  get
\begin{align*}
  \widetilde{\map Z}^{(0)} ( |A\>\<A| \otimes |B\>\<B|) = |A B\>\< AB|
  \qquad \forall A, B \in \Lin (\mathbb C^2).
\end{align*}
The same argument can be repeated for the map $ \widetilde{\map
  Z}^{(1)}$, for which we find
\begin{align*}
  \widetilde{\map Z}^{(1)} ( |A\>\<A| \otimes |B\>\<B|) = |BA\>\< BA|
  \qquad \forall A, B \in \Lin (\mathbb C^{2}).
\end{align*}
Note that the above equations, along with linearity, define uniquely
the maps $ \widetilde{\map Z}^{(0)} $ and $ \widetilde{\map Z}^{(1)}
$.  From these facts we derive the following conclusions: \emph{i)}
there exists only one supermap on no-signalling channels that  satisfies Eq.
(\ref{suiccio}), and \emph{ii)} Eq. (\ref{suiccio}) must hold not only
for quantum channels $\map A \in \Chan (\sH_\rA \to \sH_\rA)$ and $\map B \in \Chan (\sH_\rB \to \sH_\rB)$ , but also
for arbitrary quantum operations  $\map Q_A \in \QO (\sH_\rA \to \sH_\rA)$ and $\map Q_B \in \QO (\sH_\rB \to \sH_\rB)$
This concludes the proof. $\blacksquare$

\medskip 

{\bf Remark  (impossibility of switching boxes in dimension $d>2$)} 
The impossibility proof uses the properties of Pauli matrices. With a little amount of extra labour, using the property of the shift-and-multiply unitaries it is possible to show that the same impossibility proof holds for the switch supermap defined on pair of channels in general dimension $d>2$.

%Finally, we note that the map $\map Z$ defined in Eq. (\ref{zeta}) and, more generally, any supermap $\map Z$ sending no-signalling channels in  $\Chan (\rA \rB \to \rA' \rB')$ to channels in $\Chan (\rC \to \rC')$ can be equivalently represented as a higher-order map transforming channels in  $\Chan (\rA \to \rA')$ into supermaps of type $(\rB \to \rB')  \to (\rC \to \rC')$  %(generalization 1 of Def. \ref{def:supermap}).  Indeed, we can define the higher-order map $\widehat{\map Z}$ in the %following way:  for every channel $\map A \in  \Chan (\rA\to \rA')$, the supermap $\widetilde {\map Z}  (\map A)$ is given by 
%\begin{align}
%[\widetilde {\map Z}(\map A)  ] (\map B)  :  =  \map Z  (\map A \otimes \map B)  
%\qquad \forall \map B \in  \Chan (\rB \to \rB'). 
%\end{align}  

\section{No go theorem for the classical switch of black boxes}\label{sec:noswitch}  
%In most quantum algorithms the input data are encoded in the unitary
%transformation performed by a black box and the implementation of the
%algorithm consists in the evolution of qubits through a quantum
%circuit which simply \emph{contains} the available black boxes as
%elements.  Is this a general rule?  Do quantum circuits allow for the
%computation of all possible functions whose input is a black box,
%rather than a qubit register? 
% These questions are inspired by
%Church's notion of computation \cite{barend}, which allows one to
%compute functions of functions, rather than only functions of bits.

As anticipated in the previous sections, we will now show that there exist functions of black boxes that are implementable by means of elementary
operations, but cannot be represented by a circuit obeying rules
1-4.

The key counterexample 
is provided by the switch supermap, which corresponds to the following function of two qubit black boxes $\boxed f$ and $\boxed g$ and of a classical control
bit $x$:
\begin{equation}
{\tt SWITCH}\left(x,\boxed{f},\boxed{g}\right)=\left\{
  \begin{aligned}
    \Qcircuit @C=1em @R=.7em @! R { & \gate{f} &\gate{g}&\qw}&&x=1\\
    \Qcircuit @C=1em @R=.7em @! R { & \gate{g} &\gate{f}&\qw}&&x=0
  \end{aligned}
\right.
\label{funcswitch}
\end{equation}
The two black boxes $\boxed{f}$ and $\boxed{g}$---along with the classical
bit $x$---are the \emph{input} of the function, and must be regarded
as {\em single} calls to two different oracles during the computation.  The
above example can be generalized in various ways, for example by
putting between $f$ and $g$ a third box $\boxed {U_x}$ that depends on
the value of the bit $x$, or by leaving between $f$ and $g$ an open
slot in which a third arbitrary transformation can be inserted. 

It is easy to imagine a physical device that implements the function
{\tt SWITCH}.  Consider a machine with two slots, in which the user can
plug two {\em variable} boxes $\boxed{f}$ and $\boxed{g}$ at his
choice, as in the following Fig.~\ref{fig:box}.

\begin{figure}[ht]
  \epsfig{file=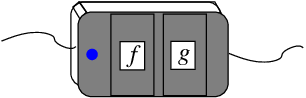,width=8cm}
  \caption{A sketch of the ideal machine implementing the {\tt SWITCH}
    function on the input boxes $\boxed{f}$ and $\boxed{g}$. \label{fig:box}}
\end{figure}

\noindent The machine is programmed with the following code:

\noindent
{\tt\obeylines
PROGRAM "SWITCH"
if  $x=1$
\quad then 
\qquad do $\Qcircuit @C=1em @R=.7em @! R { & \gate{f} &\gate{g}&\qw}$
\quad else 
\qquad do $\Qcircuit @C=1em @R=.7em @! R { & \gate{g} &\gate{f}&\qw}$
endif}

\bigskip 

We can imagine that the machine has movable wires inside, that can
connect the boxes $\boxed f$ and $\boxed g$ in two possible ways
depending on the value of the classical bit $x$, thus implementing the
{\tt SWITCH} function.  Ordinary quantum circuits, however, do not
have such movable wires. They can have controlled swap operations, but
once a time-ordering between $\boxed f$ and $\boxed g$ has been chosen
in the circuit, there is no way to reverse it. Intuitively, if $g$ has
been applied after $f$, the only way to invert the order is to send
information back in time, using a fictional time machine.  We will now
make this statement rigorous, proving that if one could implement the
{\tt SWITCH} function by inserting the boxes $\boxed f$ and $\boxed g$
in a quantum circuit, then the same circuit could be used to implement
deterministic time-travel.  Since deterministic time travel is
impossible in standard quantum mechanics, this fact leads to the
following no-go theorem.

\begin{theorem}[No classical switch of boxes]\label{th:noswitch}
  The function {\tt SWITCH}  defined in Eq.   (\ref{funcswitch})  cannot be computed deterministically by a
   circuit in which the two unknown oracles
  $\boxed{f}$ and $\boxed{g}$ are called a single time in a fixed causal order.
\end{theorem}

As anticipated, the proof is by contradiction: we will now prove that if the function {\tt SWITCH} could be implemented by inserting the boxes in a circuit, then that circuit could be used to send qubits back in time.      

\begin{proposition}{\bf (Switching boxes in a circuit implies the deterministic time travel)}\label{prop:timetravel}
  If the function {\tt SWITCH}  defined in Eq.   (\ref{funcswitch})  could be implemented on an arbitrary pair of black boxes
  $\boxed{f}$ and $\boxed{g}$ by inserting $\boxed{f}$ and $\boxed{g}$ in a circuit, then the same circuit could be used to achieve deterministic time travel.  
\end{proposition}

{\bf Proof.} Suppose by absurd that there exists a deterministic
circuit performing the program {\tt SWITCH} using a single call to
$\boxed f$ and $\boxed g$.  Without loss of generality, let us assume
that in this circuit the oracle $\boxed{f}$ is called before the
oracle $\boxed{g}$.  Then we must have 
\begin{align}
  \nonumber 
  &\begin{aligned}\Qcircuit @C=1em @R=.7em @! R {& & \multigate{1}{\map C_1}&\gate{f}
      &\multigate{1}{\map C_2}&\gate{g}&\multigate{1}{\map C_3}& \qw \\
      |x \rangle\langle x|  ~~  & & \ghost{\map C_1}&\qw &\ghost{\map C_2}&\qw &\ghost{\map C_3}&\\
      }
  \end{aligned}  \\
  & \qquad\qquad \qquad \quad =
  \left\{ \begin{aligned}\Qcircuit @C=1em @R=.7em @! R {&\gate{f} & \gate g &\qw& & x=1\\
        &\gate g &\gate f &\qw && x=0}
    \end{aligned}\right.  \label{eq:circswitch}
\end{align}
where $\map C_1$, $\map C_2$ and $\map C_3$ are quantum channels (possibly
using ancillary systems).

Now, let $\map S : \Herm(\rA\rB\to\rA'\rB')\to \Herm(\rA\rQ\to\rA)$ be the linear map defined
by the above circuit, namely the linear map defined by 
\begin{align*}
\map S ( \map A \otimes \map B)  :  =    \map C_3 (  \map B \otimes \map I_3 )    \map C_2  (  \map A \otimes \map I_1 ) \map C_1
\end{align*}
where $  \map A  \in \Herm (\rA \to \rA')$ and $  \map B  \in \Herm (\rB \to \rB')$ are generic maps  and $\map I_1$ and $\map I_2$ denote the identity on the ancillary qubits at steps 1 and 2, respectively, so that for all
channels $\map A,\map B$ it holds that the channel depicted in Eq.~\eqref{eq:circswitch}  is given by
$\map S(\map A \otimes \map B)$.

By definition, $\map S$ is a supermap on product channels: it sends product channels to quantum channels, even when acting on bipartite product channels (see definition \ref{def:supermapprod}).   Since the set of supermaps on product channels  coincides with the set of supermaps on no-signalling channels (theorem \ref{theo:alternative}), $\map S$ is also a map on no-signalling channels. Moreover, by hypothesis [eq. (\ref{eq:circswitch})] $\map Z$ satisfies Eq. (\ref{zeta}).  Hence, $\map S$ is
exactly the supermap $\map Z$ defined in subsection \ref{subsect:switch}.  

Now, by lemma \ref{lem:switchext} we know that Eq. (\ref{eq:circswitch})   must hold also when $f$ and $g$ are arbitrary quantum operations. We will now show that this leads to a contradiction.  Let us introduce an additional qubit  $\rE$.   Now,   every bipartite channel $\map   F \in \Chan (\rA \rE  \to \rA' \rE)$  can be written as a linear combination $  \map F  =  \sum_{i,j}    x_{ij}   ~   f_i  \otimes e_j $, where each $x_{ij}$ is a (possibly negative) real number,  $ f_i\in \QO (\rA \to \rA')$ and $e_j  \in  \QO (\rE \to \rE)$ are suitable quantum operations, and similarly  every bipartite channel $\map   G \in \Chan (\rB \rE  \to \rB' \rE)$ can be written as   $  \map G  =  \sum_{kl }    y_{kl}   ~   g_k  \otimes e_l$, with suitable coefficients $y_{kl}$ and suitable quantum operations $g_k  \in  \QO (\rB \to \rB')$.    Hence, by linearity, we obtain that for $x=0$ the fixed circuit  locally switches
bipartite boxes, that is, we have for generic two-qubit channels $\map
F$ and $\map G$
\begin{align}
\nonumber &  \begin{aligned}\Qcircuit @C=1em @R=.7em @! R {
      & &\qw&\ghost{\map F}  &\qw   &      &    \ghost{\map G} &\qw&\qw\\
      & & \multigate{1}{\map C_1}&\multigate{-1}{\map F} &\multigate{1}{\map C_2}&\qw &\multigate{-1}{\map G}&\multigate{1}{\map C_3}& \qw\\
      |x \rangle \langle  x|  ~~& & \ghost{\map C_1}&\qw &\ghost{\map C_2}&\qw  &\qw &\ghost{\map C_3}&   }
  \end{aligned} \\  
 \nonumber &  \\
  \label{bipartite switch}
 &  \qquad \qquad   \qquad \qquad =
   \left\{ \begin{aligned}
      &\Qcircuit @C=1em @R=.7em @! R {&\multigate{1}{\map F} &  \qw & &  \multigate{1}{\map G} &\qw&\\
        & \ghost{\map F}  & \qw&\qw & \ghost{\map G}  &\qw }&x=1\\
      &\begin{aligned} \mbox{ \setlength{\unitlength}{.01pt}
          \begin{picture}(5288,3324)(1200,-5173)
            {\put(2851,-2161){\line( 1, 0){600}} }%
            {\put(4051,-2161){\line( 1, 0){600}} }%
            \put(1886,-2911){\makebox(0,0)[lb]{\smash{{$\map F$}}}}
            {\put(4651,-3661){\framebox(1200,1800){}} }%
            \put(5026,-2911){\makebox(0,0)[lb]{\smash{{$\map G$}}}}
            {\put(3076,-3736){\oval(750,750)[tr]} }%
            {\put(4326,-3736){\oval(750,750)[tl]} }%
            {\put(4126,-4456){\oval(1400,1400)[bl]} }%
            {\put(3276,-4456){\oval(1400,1400)[br]} }%
            {\put(6451,-3961){\oval(900,1200)[br]}
              \put(6451,-3961){\oval(900,1200)[tr]} }%
            {\put(1051,-3961){\oval(900,1200)[tl]}
              \put(1051,-3961){\oval(900,1200)[bl]} }%
            {\put(5851,-2161){\line( 1, 0){750}} }%
            {\put(901,-2161){\line( 1, 0){750}} }%
            {\put(2851,-3361){\line( 1, 0){325}} }%
            {\put(4326,-3361){\line( 1, 0){325}} }%
            {\put(3426,-3736){\line( 0,-1){720}} }%
            {\put(4126,-5161){\line( 1, 0){2250}} }%
            {\put(3976,-3736){\line( 0,-1){720}} }%
            {\put(1026,-5161){\line( 1, 0){2250}} }%
            {\put(6551,-4561){\line(-1, 0){5600}} }%
            {\put(1051,-3361){\line( 1, 0){600}} }%
            {\put(5851,-3361){\line( 1, 0){600}} }%
            {\put(1651,-3661){\framebox(1200,1800){}}}%
          \end{picture}}
      \end{aligned}& x=0
    \end{aligned}\right.   
\end{align}
where the backward line in the $x=0$ case  is a graphical notation meaning that the second output of  channel $\map G$ is fed in the second input of channel $\map F$. 

Now consider the case of two swap channels $\map F = \map G = \map E$,
with $\map E (\rho \otimes \sigma) = \sigma\otimes \rho$.  In this
case, the output for $x=0$ would be a circuit containing a time loop,
as represented in the following diagram:
\begin{align}
\nonumber &  \begin{aligned}\Qcircuit @C=1em @R=.7em @! R {
      &\ustick{\rA_1}\qw&\qw&\ghost{\map E}  &\ustick{\rA_2}\qw   &\qw&&\ustick{\rA_3}\qw& \ghost{\map E} &\qw&\ustick{\rA_4}\qw\\
      &\ustick{\rB_1}\qw& \multigate{1}{\map C_1}&\multigate{-1}{\map E} &\qw&\multigate{1}{\map C_2}&\qw&\qw&\multigate{-1}{\map E}&\multigate{1}{\map C_3}& \ustick{\rB_2}\qw\\
      |0 \rangle \langle  0|  ~~&& \ghost{\map C_1}&\qw&\qw &\ghost{\map C_2}&\qw&\qw&\qw&\ghost{\map C_3}&   }
  \end{aligned} =  \\  
 \nonumber &  \\  
\nonumber &  \qquad \qquad \qquad \qquad  =\quad
%\begin{aligned}
%\mbox{\epsfig{file=noswitch.eps,width=3.5cm}}
%\end{aligned}  \\
\begin{aligned} 
      \mbox{\setlength{\unitlength}{.01pt}
        \begin{picture}(8288,3324)(1200,-5173)
          \put(3051,-1861){\makebox(0,0)[lb]{\smash{{$\rA_2$}}}}
          {\put(2851,-2161){\line( 1, 0){1200}} }%
         \put(4851,-1861){\makebox(0,0)[lb]{\smash{{$\rA_3$}}}}
           {\put(5051,-2161){\line( 1, 0){1200}} }%
          \put(2026,-2911){\makebox(0,0)[lb]{\smash{{$\map E$}}}}
          {\put(6251,-3661){\framebox(1200,1800){}} }%
          \put(6526,-2911){\makebox(0,0)[lb]{\smash{{$\map E$}}}}
          {\put(3876,-3736){\oval(750,750)[tr]} }%
          {\put(5126,-3736){\oval(750,750)[tl]} }%
          {\put(4926,-4456){\oval(1400,1400)[bl]} }%
          {\put(4076,-4456){\oval(1400,1400)[br]} }%
          {\put(8151,-3961){\oval(900,1200)[br]}
            \put(8151,-3961){\oval(900,1200)[tr]} }%
          {\put(1051,-3961){\oval(900,1200)[tl]}
            \put(1051,-3961){\oval(900,1200)[bl]} }%
         \put(7751,-1861){\makebox(0,0)[lb]{\smash{{$\rA_4$}}}}
          {\put(7551,-2161){\line( 1, 0){1200}} }%
          \put(351,-1861){\makebox(0,0)[lb]{\smash{{$\rA_1$}}}}
          {\put(451,-2161){\line( 1, 0){1200}} }%
          {\put(2851,-3361){\line( 1, 0){1125}} }%
          {\put(5126,-3361){\line( 1, 0){1125}} }%
          {\put(4226,-3736){\line( 0,-1){720}} }%
          {\put(4926,-5161){\line( 1, 0){3250}} }%
          {\put(4776,-3736){\line( 0,-1){720}} }%
          {\put(826,-5161){\line( 1, 0){3250}} }%
          {\put(8151,-4561){\line(-1, 0){7200}} }%
          {\put(1051,-3361){\line( 1, 0){600}} }%
          {\put(7551,-3361){\line( 1, 0){600}} }%
          {\put(1651,-3661){\framebox(1200,1800){}}}%
          \put(351,-6161){\makebox(0,0)[lb]{\smash{{$\rB_1$}}}}
          \put(7751,-6161){\makebox(0,0)[lb]{\smash{{$\rB_2$}}}}
          \end{picture}}
      \end{aligned}\quad 
  \nonumber &  \\  
\nonumber\\
  &  \qquad \qquad \qquad \qquad=\qquad\qquad
\begin{aligned} 
  &\mbox{\setlength{\unitlength}{.01pt}
    \begin{picture}(5288,3324)(1200,-5173) {
        {\put(7451,-3961){\oval(1200,2200)[br]}
          \put(7451,-3961){\oval(1200,2200)[tr]}}
        \put(0051,-4761){\makebox(0,0)[lb]{\smash{{$\rA_2$}}}}
        \put(0051,-5061){\line(1,0){1000}}%
%        \put(0151,-5061){\line(1,0){200}}%
        \put(0051,-2861){\line(1,0){7400}}%
        \put(6651,-4761){\makebox(0,0)[lb]{\smash{{$\rA_3$}}}}
        \put(6651,-5061){\line(1,0){800}}%
        {\put(0051,-3961){\oval(1200,2200)[tl]}
          \put(0051,-3961){\oval(1200,2200)[bl]} }}
    \end{picture}}\\
 \\
  &\Qcircuit @C=1em @R=.7em @! R { 
      &\ustick{\rA_1}\qw&  \multigate{1}{\map E}  &  \qw & \ustick{\rA_4}\qw\\
     &\ustick{\rB_1}\qw&  \ghost{\map E} & \qw &\ustick{\rB_2}\qw}   
        \end{aligned}
\label{backfromfuture} 
\end{align}
where the last equality can be easily verified considering that the
swap gate $\map E$ acts as an identity map from the top left system to
the bottom right, and as an identity from the bottom left to the top
right. The loop on top of the swap channel represents an identity map
from a future computational step $\rA_3$ to a previous one $\rA_2$ (in
other words, a deterministic time travel).  $\blacksquare$

\medskip  Having reduced the circuit realization of the {\tt SWITCH} program to the realization of a time travel machine means having proved its impossibility.  A formal proof  is given in the following.  

{\bf Proof of theorem \ref{th:noswitch}.}  Consider  probabilistic teleportation, represented by the equation
\begin{align}\label{probtele}
  \begin{aligned}
    \Qcircuit @C=1em @R=.7em @! R {&\multiprepareC{1}{\Phi^+}&\qw&\qw&\qw\\
      &\pureghost{\Phi^+}&\qw&\multimeasureD{1}{E}\\
      & \qw &\qw&\ghost{E}}
  \end{aligned}
  = \frac 14 ~ \begin{aligned} \Qcircuit @C=1em @R=.7em @! R
    {&\gate{\map{I}}&\qw}
  \end{aligned}~~,
\end{align}
where $\Phi^+$ represents the preparation of a maximally entangled
state of two qubits, $E$ represents the outcome of the Bell
measurement corresponding to the projection on $\Phi^+$, and $\map I$
is the identity channel for a single qubit.    Multiplying both members  by $4$,  Eq. (\ref{probtele})  becomes a way to represent the identity channel.   For an identity channel from the future to the past, we have   
\begin{align*}
\qquad\quad
\begin{aligned} 
  &\mbox{\setlength{\unitlength}{.01pt}
    \begin{picture}(5288,3324)(1200,-5173) {
        {\put(7451,-3961){\oval(1200,2200)[br]}
          \put(7451,-3961){\oval(1200,2200)[tr]}}
        \put(0051,-2861){\line(1,0){7400}}%
        \put(0051,-5061){\line(1,0){200}}%
        \put(7251,-5061){\line(1,0){200}}%
        {\put(0051,-3961){\oval(1200,2200)[tl]}
          \put(0051,-3961){\oval(1200,2200)[bl]} }}
    \end{picture}}   
        \end{aligned}  \quad
=    4  \begin{aligned}
    \Qcircuit @C=1em @R=.7em @! R {&\multiprepareC{1}{\Phi^+}&\qw&   \qw&\multimeasureD{1}{E} \\
      &\pureghost{\Phi^+}&\qw &  &  \ghost{E} }
  \end{aligned}
\end{align*}  

Substituting this identity in Eq. (\ref{backfromfuture}), we obtain
\begin{align*}
\nonumber &  \begin{aligned}\Qcircuit @C=1em @R=.7em @! R {
      & &\qw&\ghost{\map E}  &\qw   &      &    \ghost{\map E} &\qw&\qw\\
      & & \multigate{1}{\map C_1}&\multigate{-1}{\map E} &\multigate{1}{\map C_2}&\qw &\multigate{-1}{\map E}&\multigate{1}{\map C_3}& \qw\\
      |0 \rangle \langle  0|  ~~& & \ghost{\map C_1}&\qw &\ghost{\map C_2}&\qw  &\qw &\ghost{\map C_3}&   }
  \end{aligned} =  \\  
 &\\
 &  \qquad\qquad\qquad \qquad    = 4 ~ ~  \begin{aligned}\Qcircuit @C=1em @R=.7em @! R { \multiprepareC{1}{\Phi^+}&\qw& \qw &  \qw&\multimeasureD{1}{E} \\
       \pureghost{\Phi^+}&\qw &   &&  \ghost{E} \\
      &\qw&  \multigate{1}{\map E}  &  \qw & \qw\\
     &\qw&  \ghost{\map E} & \qw &\qw}
  \end{aligned}
\end{align*}
Finally, connecting the top wires gives
\begin{align*}
\nonumber &  \begin{aligned}\Qcircuit @C=1em @R=.7em @! R {
      & &\qw&\ghost{\map E}  &\qw   &  \qw    &    \ghost{\map E} &\qw&\qw\\
      & & \multigate{1}{\map C_1}&\multigate{-1}{\map E} &\multigate{1}{\map C_2}&\qw &\multigate{-1}{\map E}&\multigate{1}{\map C_3}& \qw\\
      |0 \rangle \langle  0|  ~~& & \ghost{\map C_1}&\qw &\ghost{\map C_2}&\qw  &\qw &\ghost{\map C_3}&   }
  \end{aligned} =  \\  
 &\\
 &  \qquad \qquad    = 4 ~ ~  \begin{aligned}\Qcircuit @C=1em @R=.7em @! R { \multiprepareC{1}{\Phi^+}&\qw& \qw &  \qw&\multimeasureD{1}{E} \\
       \pureghost{\Phi^+}&\qw & \qw  &\qw &  \ghost{E} \\
      &\qw&  \multigate{1}{\map E}  &  \qw & \qw\\
     &\qw&  \ghost{\map E} & \qw &\qw}
  \end{aligned}\\
  &\\
  &  \qquad \qquad    = 4 ~ ~  \begin{aligned}\Qcircuit @C=1em @R=.7em @! R { 
      &\qw&  \multigate{1}{\map E}  &  \qw & \qw\\
     &\qw&  \ghost{\map E} & \qw &\qw}
  \end{aligned}
\end{align*}
This is clearly absurd because the first term in the chain of equalities it is trace-preserving, while the last term is not.   In fact, the above equation implies the absurd statement $1=4$.
$\blacksquare$

\medskip 

{\bf Remark 1 (Impossible switches and impossible time-travels).}  As we saw in proposition \ref{prop:timetravel}, a circuit switching black boxes would enable  a \emph{deterministic time-travel}, where the state of a  qubit on the top is teleported back into the past.  It is worth mentioning that the converse is also true:   having access to an hypothetical time travel machine sending qubits from the future to the past would allow one to build a
computational circuit for the program {\tt SWITCH}.     As in the proof of proposition  \ref{prop:timetravel}, we will represent the time travel machine by a probabilistic teleportation diagram, suitably rescaled by a factor 4 (cf.  Eq. (\ref{backfromfuture}), following the model of closed time-like curves considered  in Refs. \cite{benschu,coecke,svetli,MIT}.   It is
known that such an artificial rescaling of the probability of postselected outcomes has dramatic computational consequences \cite{aaronson}. In our case, it would allow one to construct a circuit that realizes the {\tt SWITCH} transformation. 

\begin{proposition}\label{prop:travelswitch}{\bf (Closed timelike curves enable a circuit realization of the {\tt SWITCH} program)}   If  access to a closed timelike curve were available, then the program
${\tt SWITCH}$ could be implemented deterministically by inserting the two black boxes $f$ and $g$ in a circuit.
\end{proposition}

{\bf Proof.}   It is immediate to check the  equality
\begin{align*}
{\tt SWITCH}\left(x,\boxed{f},\boxed{g}\right)   =   4 \begin{aligned}
  \Qcircuit @C=1em @R=.7em @! R {&&\ctrl{1}&\gate{\map X}&\ctrl{1}&\measureD{  \Tr}\\
  &&\multigate{1}{\map E}&\gate{g}&\multigate{1}{\map E}&\qw\\
  &\multiprepareC{1}{\Phi^+}&\ghost{\map E}&\gate{f}&\ghost{\map E}&\multimeasureD{1}{E}\\
  &\pureghost{\Phi^+}& \qw &\qw &\qw&\ghost{E}&}
  \end{aligned}
  \end{align*} 
where   $\Tr$ represents the partial trace,   $\map X$ is the bit-flip channel $\map X  (  \rho )  =  X\rho X$, $ X  =  |0\rangle\langle  1| +  |1\rangle\langle  0|  $ and $\begin{aligned}
  \displaystyle\Qcircuit @C=.5em @R=0em @!R {&\qw &\ctrl{1}&\qw\\ &\qw&\multigate{1}{\map E}&\qw\\
    &\qw&\ghost{\map E}&\qw} \end{aligned}$ represents the control-SWAP channel  $\map E  (\rho)  =  U \rho U^\dag$, $U  =  I\otimes  |0\rangle \langle 0|  +  SWAP  \otimes  |1\rangle \langle 1|    $,   $  SWAP  |\alpha\rangle|\beta\rangle  =  |\beta\rangle |\alpha\rangle$.      
$\blacksquare$  

\medskip 

Combining propositions  \ref{prop:timetravel} and \ref{prop:travelswitch}, we then obtain the following equivalence:   
\begin{corollary}{\bf (Switching boxes in a circuit is equivalent to time travel)}
The program ${\tt SWITCH}$ can be implemented deterministically by inserting the two black boxes $f$ and $g$ in a circuit if and only if access to a closed timelike curve is available.
\end{corollary}

\medskip

{\bf Remark 2 (relation with Church's $\lambda$-calculus).} The program {\tt SWITCH} is the prototype of a {\em higher-order
  computation} of the kind described in the $\lambda$-calculus by
Church \cite{barend}.  In this model, the input and output of a computation can be  functions, instead of blocks
of data.   Theorem 1 states that there exists an  higher-order computation that cannot be implemented by a
quantum circuit  containing only one use of $\boxed f$ and $\boxed
g$ in a pre-defined causal order.   

The idea to construct a formal language able to encode a quantum version of Church's $\lambda$-calculus has been considered by several authors in the literature, leading to many different versions of quantum $\lambda$-calculi \cite{vanTonder,selinger,grattage,sv2004,perdrix,arrighi}. 
It is interesting to note that the program  {\tt SWITCH}  is an example of the computations that can be expressed in the version by Selinger and Valiron \cite{sv2004} of a $\lambda$-calculus  for quantum computations with classical control.  
Later in the paper we will also consider the quantum version of the program {\tt SWITCH}, which is an example of higher-order computation outside the model of Ref. \cite{sv2004}. 
   
\medskip 

{\bf Remark 3 (Impossibility of switching classical boxes).}  The impossibility of implementing the program {\tt SWITCH} by insertion of the input boxes in a computational circuit obeying rules 1-4 holds not only in the
quantum world, but also in the classical one.
Indeed, the proof given in the quantum case can be adapted to the classical case by substituting Eq. (\ref{probtele})  with the diagram for classical probabilistic teleportation using a  maximally correlated mixed state. 
The impossibility of a circuit realization of the {\tt SWITCH} program is a very basic fact, and as such might have been observed in the literature in classical computer science.  However, to the best of our knowledge, Theorem \ref{th:noswitch} is the first actual proof of it.

\section{Ways around the no-go theorem}\label{sec:waysaround}  
The problem with the realization of  the program {\tt SWITCH} by insertion in a
ordinary circuit is due to four different facts that are assumed in the hypothesis of the no-go theorem:
\begin{enumerate}
\item the facts that the functions $f$ and $g$ are  provided as \emph{black boxes}   
\item the fact that the black boxes can be called \emph{only once} in the run of the circuit
\item the fact that  \emph{time loops are forbidden}
\item the fact that the circuit is required to be \emph{deterministic}.        
\end{enumerate}

We will now show that, by relaxing any of these requirements, one can find a way around the no-go theorem of the previous section.  

\subsection{Implementation of the program {\tt SWITCH} via access to program states}  
The first reason for the impossibility of implementing the function SWITCH problem arises  from the fact that the
input functions $f$ and $g$ are provided as physical machines (black boxes)
inserted in a circuit.  This problem would not arise if the functions $f$ and $g$ were encoded into sets of programming data defining two subroutines. Indeed, when functions are encoded into
strings of (qu)bits, they can be processed sequentially by a circuit  using controlled operations.  More precisely, suppose that we are given two  \emph{program states}  $\rho_f, \rho_g \in \St(\rP)$  ($\rP$ being the  program system) and a programmable channel  $\map R  \in  \Chan  (\rA  \rP  \to \rA)$ such that
\begin{align*}
 \begin{aligned}
    \Qcircuit @C=1em @R=.7em @! R 
    {    & &  \qw \poloFantasmaCn{\rA}  &  \multigate{1}{  \map R}  & \qw\poloFantasmaCn{\rA}  &\qw  \\
      \rho_f  &  &  \qw \poloFantasmaCn{\rP}   &  \ghost{\map R}   &    & }
  \end{aligned}
 & =
  \begin{aligned}  \Qcircuit @C=1em @R=.7em @! R
    {&\gate{f}&\qw}
  \end{aligned}\\ 
  &\\
  \begin{aligned}
  \Qcircuit @C=1em @R=.7em @! R 
    {    & &  \qw \poloFantasmaCn{\rA}  &  \multigate{1}{  \map R}  & \qw\poloFantasmaCn{\rA}  &\qw  \\
      \rho_g &  &  \qw \poloFantasmaCn{\rP}   &  \ghost{\map R}   &    & }
  \end{aligned}
 & =
  \begin{aligned}  \Qcircuit @C=1em @R=.7em @! R
    {&\gate{g}&\qw}
  \end{aligned}.
\end{align*}

In that case, the output of the program {\tt SWITCH} for the particular input pair $(\boxed f ,  \boxed g)$can be produced as follows
\begin{align*}
{\tt SWITCH}\left(x,\boxed{f},\boxed{g}\right)   =  \quad
 \begin{aligned}
  \Qcircuit @C=1em @R=.7em @! R 
    {                       &&                   \qw                    & \multigate{1}{  \map R}  &  \multigate{2}{\map R}  &\qw  \\
      \rho_g    &&    \ghost{\map E}           &    \ghost{\map R}   &     \pureghost{\map R}                 & \\
      \rho_f & & \multigate{-1}{\map E}   &             \qw                    &  \ghost{\map R} &          \\
                & &  \ctrl{-1}                        &     \qw & \qw & \measureD{\Tr} }
  \end{aligned}
  \end{align*}

However, such a realization is possible only for those black boxes $\boxed f$ and $\boxed g$ that can be encoded in the state of the program system and decoded by a programmable channel $\map R$.  In quantum theory, the
no-programming theorem \cite{nc} states that it is impossible  to encode an arbitrary quantum channel in the state of a finite quantum system.  This is due to the fact that two unitary channels can be retrieved from their program states if and only if  the program states are orthogonal. 
   
\subsection{Implementation of the {\tt SWITCH} program with two queries to the black boxes}  

Another obstacle to the realization of the {\tt SWITCH} program arises from the fact that the oracles $f$ and $g$ are
restricted to be called \emph{only once}, i.e.  that the circuit must contain
boxes $\boxed{f}$ and $\boxed g$ only once (rule 4) and in a definite
time order (rule 3).  Indeed, a computational circuit that produces
the \emph{same output} of the program {\tt SWITCH} actually exists,
but it requires two calls to at least one of the oracles $f$ and $g$,
e.~g. as follows
\begin{equation}\label{condueg}
  \begin{aligned}\Qcircuit @C=1em @R=.7em @! R {|x\rangle &  &\qw & \ctrl{1}  &\qw & \ctrl{1} &\gate{\map X}&\ctrl{1}&\qw & \ctrl{1} &  \qw \\
      &   &\qw & \multigate{1}{\map E}   &\gate{g} &  \multigate{1}{\map E}  &   \gate{f} &\multigate{1}{\map E}  &   \gate{g}  & \multigate{1}{\map E}& \qw\\
      &  &\qw & \ghost{\map E}  &\qw &\ghost{\map E} &\qw&\ghost{\map E}&\qw & \ghost{\map E} & \qw}
 \end{aligned}
  \end{equation}
where $\begin{aligned}
  \displaystyle\Qcircuit @C=.5em @R=0em @!R {&\qw &\ctrl{1}&\qw\\ &\qw&\multigate{1}{\map E}&\qw\\
    &\qw&\ghost{\map E}&\qw} \end{aligned}$ is a control-swap channel,
exchanging the two input qubits depending on the state of the control
qubit, and $\boxed{\map X}$ is the bit flip channel. 
The above circuit achieves the desired ${\tt SWITCH}$ transformation over the qubit in the
middle wire depending on the state of the controlling qubit at the top
wire. 
This fact is \emph{not} in contradiction with Theorem 1: If the input are two black boxes $\boxed f$,
$\boxed g$, the possibility of achieving two uses from a single one is
ruled out by the no-cloning theorem for boxes \cite{ourclon}. Again,
the limitation due to the single call constraint is strictly related
to the black box  nature of the functions $f$ and $g$. If we knew what $f$ and $g$ are, we would be  duplicate them, thus making possible the computation of the function
$\map S(x,\boxed{f},\boxed{g})$ through the circuit of Eq. (\ref{condueg}).

\subsection{Implementation of the program {\tt SWITCH} through access to a closed timelike curve}

This point was already discussed in proposition \ref{prop:travelswitch}: a circuit that has access to a closed timelike curve (i.e. an identity channel from the future to the past) can implement the program {\tt SWITCH} deterministically, on arbitrary black boxes, by running the black boxes only once.

\subsection{Probabilistic simulation of the {\tt SWITCH} program with a single query to the black boxes }  
Another factor that prevents the implementation of the program {\tt
  SWITCH} as a computational circuit is the requirement that the
program succeeds \emph{deterministically}. Indeed, rules 1-4 do not forbid
achieving the task with some probability. In particular, a
computational circuit that uses probabilistic teleportation succeeds
in the task with probability $1/4$ is given by 
\begin{equation*}
  \Qcircuit @C=1em @R=.7em @! R {&&\ctrl{1}&\gate{X}&\ctrl{1}&\measureD{\Tr}\\
  &&\multigate{1}{\map E}&\gate{g}&\multigate{1}{\map E}&\qw\\
  &\multiprepareC{1}{\Phi^+}&\ghost{\map E}&\gate{f}&\ghost{\map E}&\multimeasureD{1}{E}\\
  &\pureghost{\Phi^+}& \qw &\qw &\qw&\ghost{E}&}
\end{equation*}
When the outcome $E$ occurs in this circuit, we may say that the third
qubit (from the top) has been teleported from the future back to the
past.  In this case it is easy to see that if the control qubit is in
state $\ket{1}$ one obtains the sequence ``$\boxed f$ followed by
$\boxed g$'' acting on the second input qubit, while if the control
qubit is in state $\ket{0}$ the boxes are exchanged.  What's more, if
one puts the control qubit in the superposition $(\ket{0}+\ket{1})/\sqrt 2$ and omits the partial trace $\Qcircuit @C=1em @R=.7em @! R {&\measureD{\Tr}}$, one
 obtains a quantum superposition of the two orderings of the boxes, namely
the output of the circuit is proportional to
$(U_fU_g\ket{\psi}\ket{1}+U_{g}U_{f}\ket{\psi}\ket{0}  )/\sqrt 2$, where
$\ket{\psi}$ is the input state of the qubit in the second wire, and
$U_f$ and $U_g$ denote the unitary operators corresponding to boxes
$\boxed{f}$ and $\boxed{g}$, respectively. Note, however, that the
probability of achieving the program {\tt SWITCH} for $\boxed f$ and
$\boxed g$ transforming $N$ qubits goes to zero exponentially as
$4^{-N}$ versus the number $N$ of input qubits for each box.  The probability $p_N  =  4^{-N}$ is actually the \emph{maximum} probability that can be achieved in a probabilistic simulation of the program {\tt SWITCH}: indeed,  proposition \ref{prop:timetravel} implies that any probabilistic simulation of the program {\tt SWITCH} with a single query to $f$ and $g$ would necessarily be a probabilistic simulation of an identity channel from the future to the past. On the other hand, Ref. \cite{genkina} shows that the maximum probability of simulating such an identity channel for $N$ qubits is $4^{-N}$.

\section{Re-modelling of the oracles in order to allow for the classical switch}\label{sec:remodelling}  
What rule in the theory of
computational circuits can be modified in order to recover the
physical implementation of the function $S(x,\boxed{f},\boxed{g})$ of
Eq.~\eqref{funcswitch}, whose computation is achieved through the
program {\tt SWITCH}? One possibility is to modify rule 3, and to
allow for circuits containing certain time loops. However, introducing
time travels in the model seems a rather drastic solution. A more
moderate approach is to modify rule 4: In particular, we may assume
that the resource provided by a single call to each of the two
physical oracles---that would be separately described as $\boxed{f}$
and $\boxed{g}$---{\em in a causal succession that can be decided by
  the user}, is described in circuital terms as a single oracle with
classical control:
\begin{equation*}
  \Qcircuit @C=1em @R=.7em @! R {&& \multigate{1}{{f/g}} &\qw  &&\multigate{1}{{g/f}} &\qw \\
    & &  \ghost{ {f/g}} &\qw &\qw &   \ghost{  {g/f}} &}
\end{equation*}
where the wire on the bottom left denotes the control qubit, whose
general state is $|\varphi\rangle=\alpha|0\rangle+\beta|1\rangle$ with
$|\alpha|^2+|\beta|^2=1$. The input $x$ is encoded on the state
$|\varphi\rangle$ as follows: For $x=0$ we prepare
$|\varphi\rangle=|0\rangle$, for $x=1$ we prepare
$|\varphi\rangle=|1\rangle$. If the two qubits on the top lines are in the states $\rho_1$ and $\rho_2$, respectively, the action of the oracle is given by
\begin{equation} \label{oracle}
  \begin{split} 
    \mathscr{O}_{f,g}(|\varphi\rangle\langle\varphi| \otimes \rho_1 \otimes \rho_2)= &|\langle 1|\varphi\rangle|^2 ~ U_f\rho_1 U_f^\dag \otimes U_g \rho_2 U_g^\dag\\
    & + |\langle 0|\varphi\rangle|^2 ~ U_g\rho_1 U_g^\dag \otimes U_f
    \rho_2 U_f^\dag
  \end{split}
\end{equation}
This way of representing the oracle is consistent with the basic
properties that one expects for the resource, namely that it perform
two successive transformations, one being a call of the box
$\boxed{f}$ and the other a call of the box $\boxed{g}$, with the
order of such calls being controlled by the variable $x$ encoded in
the state $|\varphi\rangle$. During the time interval between the
calls to the oracle, any transformation can happen, including
evolutions transforming the first output into the second input.
Exploiting the latter representation of the oracle one can clearly
implement the program {\tt SWITCH}, just by connecting the output of
the first box with the input of the second one, and encoding the bit
$x$ in the state $|\varphi\rangle$ as follows
\begin{equation*}
  \Qcircuit @C=1em @R=.7em @! R {&& \multigate{1}{{f/g}} &\qw  &\multigate{1}{{g/f}} &\qw \\
    |\varphi\rangle&& \ghost{ {f/g}} &\qw &\ghost{  {g/f}} &}
\end{equation*}
If we assume that the oracle of Eq.~\eqref{oracle} translates the
resource provided by a single use of the physical boxes corresponding
to $\boxed{f},\boxed{g}$ with classical control of the causal
ordering, we can then consider the function $S(x,\boxed{f},\boxed{g})$
as computable by a quantum circuit exploiting this resource.

\begin{figure}[h!]
  \epsfig{file=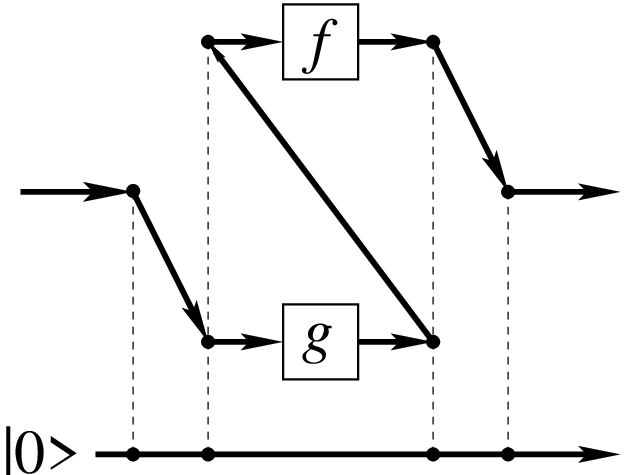,height=3cm}\quad\epsfig{file=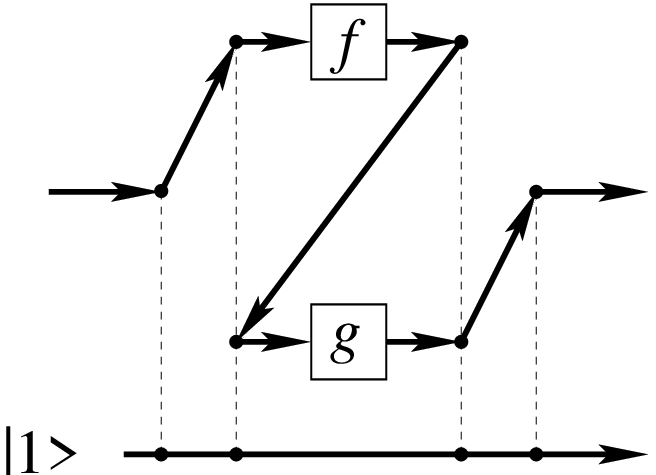,height=3cm}\\
  \caption{Quantum machine with classical control over movable
    wires.\label{aaa}}
\end{figure}

Such an oracle can be achieved in practice, for example, by a physical
circuit in which the connections between wires are movable, as in Fig.
\ref{aaa}.
Higher-order functions that transform black boxes with the assistance of classical control on the connections are described formally by the quantum $\lambda$-calculus of Ref.  \cite{sv2004}.

\section{A new resource: The quantum switch of boxes}\label{sec:quantum}   
While representing automated {\em classical control} of causal
sequences of operations  allows one to implement the program {\tt SWITCH} within the computational
circuit model, it leaves unanswered the question how {\em quantum
  control} of causal sequences of operations can be described. We can
of course imagine a further generalization of the oracle, allowing for
quantum control, with the control qubit that preserves coherence and
becomes entangled with the causal ordering of boxes $\boxed f$ and
$\boxed g$ as follows
\begin{equation*}
  \begin{aligned}
    \Qcircuit @C=1em @R=.7em @! R {&& \multigate{1}{{f/g}} &\qw  &&\multigate{1}{{g/f}} &\qw \\
      && \ghost{{f/g}} &\qw &\qw &\ghost{ {g/f}}
      &\qw}
  \end{aligned}
\end{equation*}
When $\boxed f$ and $\boxed g$  are unitary channels, the  unitary channel describing the oracle with quantum control is $\map W_{f,g} (\rho)  =  W_{f,g}  \rho  W_{f,g}^\dag$,   $W_{f,g}$ being the control unitary 
\begin{equation}
  W_{f,g}  :=|0\rangle\langle 0|\otimes U_f\otimes U_g+|1\rangle\langle 1|\otimes U_g\otimes U_f
\end{equation}
The above construction can be suitably generalized when $f$ and $g$
are not unitary boxes, but noisy quantum channels: In this case, it is
enough to use the above formula to define the Kraus operators of the
channel with quantum control in terms of the Kraus operators of the
input channels.  Precisely, if the channels $f$ and $g$ have Kraus forms $f(\rho)  =  \sum_i  f_i \rho  f_i^\dag$ and $g(\rho)  =  \sum_j  g_j \rho  g_j^\dag$, respectively, then the channel with quantum control has Kraus form 
\begin{align*}  
\map W_{f,g}  (\sigma)  =  \sum_{i,j}   W_{f_i,g_j}  \sigma  W_{f_i,g_j}^{ \dag}
\end{align*}   
with the Kraus operators $ W_{f_i,g_j}$ given by    
\begin{align*}
 W_{f_i,g_j}:
=|0\rangle\langle 0|\otimes f_i\otimes g_j +|1\rangle\langle 1|\otimes g_j\otimes f_i.
\end{align*}  
Note that the definition of the oracle $\map W_{f,g}$ is independent of the Kraus forms chosen for $f$ and $g$. The oracle with quantum control is more general and more powerful than
the classically controlled one introduced in Eq.~\eqref{oracle}.
Indeed, having $W_{f,g}$ at disposal one can implement the classically
controlled oracle $\mathcal O_{f,g}$ by using $W_{f,g}$ and then
discarding the control qubit.

How can we build the controlled oracle $\map W_{f,g}$ if we have at disposal
one use of the black boxes  $\boxed f$ and $\boxed g$?  Again, this is a question
that the circuit model is unable to answer.  In principle, there is no
physical reason to forbid the computability of the higher-order
function defined by $\map W: f \otimes g \mapsto \map W_{f,g}$. This function is
defined not only on product boxes, but also on the more general class
of \emph{non signaling} bipartite boxes, as we already discussed.      
 The function $\map W$ is linear in its argument, transforms
deterministic boxes into deterministic boxes, and can also be applied
locally to multipartite boxes without giving rise to unphysical
effects like negative probabilities. The computation of this function
is then admissible in principle.   However, although the computation of $\map W$ is
compatible with quantum mechanics, it cannot be implemented by a
circuit with the rules 1-4, due to the lack of a pre-defined causal
ordering.  Moreover, it is also possible to prove that no circuit
using the oracle with classical control $\mathcal O_{f,g}$ can
simulate the oracle with quantum control $W_{f,g}$.

\begin{figure}[h!]
  \epsfig{file=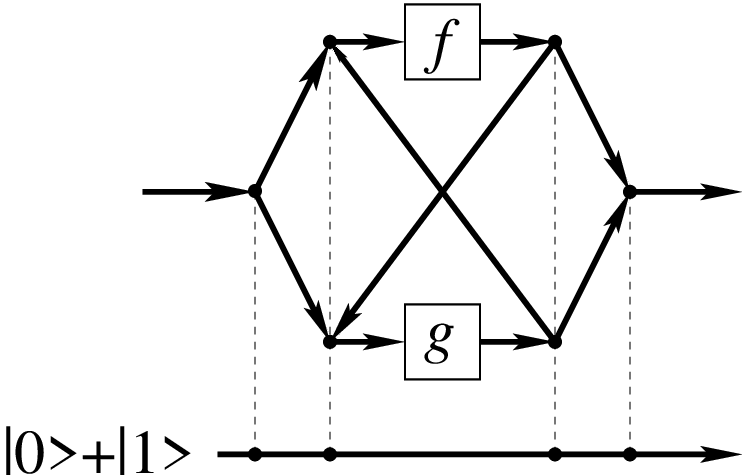,height=3cm}
  \caption{Pictorial representation of a machine with quantum control over movable
    wires.\label{bbb}}
\end{figure}

To imagine a way to build the controlled gate $W_{f,g}$ from the
boxes $\boxed f$ and $\boxed g$, we need to go beyond the usual
language of quantum circuits, and to consider also circuits with
movable wires that can be also in quantum superpositions.
For example, we can consider a thought experiment where the physical
circuit with movable wires depicted in Fig.  \ref{aaa} can be
controlled by a qubit in a way that preserves superpositions, with the
control qubit interacting with switches and controlling them in a
correlated way, as represented in Fig. \ref{bbb}. Like in the
Schr\"odinger cat thought experiment, in this case we would have a
mechanism producing entanglement between a microscopic system (the
control qubit) and a macroscopic one (the position of the switches).

%Notice however that quantum control of transformations is even more
%powerful than quantum entanglement, which is the feature giving rise
%to the typical Schr\"odinger cat state. Indeed, a
%control-unitary gate can be always used to generate a certain amount
%of entanglement. 

{\bf Remark (Simulating the quantum {\tt SWITCH} within the circuit model).}
  
The fact that the output of the quantum {\tt SWITCH} can be produced
by using two queries to the input boxes implies that a quantum circuit
model enhanced with the quantum {\tt SWITCH} is computationally
equivalent to the ordinary quantum circuit model: any oracle
computation using the quantum {\tt SWITCH} as an extra-resource can be
simulated with only a slowdown of a factor 2.  From the
complexity-theoretic point of view, the quantum {\tt SWITCH} does not
bring any extra-power in the model.  In this sense, the difference
between ordinary quantum circuits and quantum circuits powered by the
{\tt SWITCH} function is analogous to the difference between quantum
circuits and quantum Turing machines, which provide equivalent
computational models in the complexity-theoretic sense \cite{yao},
despite the fact that the simulation of a Turing machine through a
quantum circuit requires a polynomial slowdown.

Although the quantum {\tt SWITCH} can be simulated with a polynomial slowdown,  there are two important points to be made:  

\begin{enumerate}
\item  The quantum {\tt  SWITCH} does not change complexity classes, but still it offers advantages for information processing.  
For example, we may consider a problem of channel discrimination, where we have available only one use of two black boxes $  \boxed{f_i}$  and $\boxed {g_i }$, with $i=0$ or $1$, and our goal is to find out whether the label is 0 or 1.  
In these scenario, being able to implement the quantum {\tt SWITCH}  can increase the probability of successful discrimination.   For example, Ref. \cite{superseq} shows an example where the quantum {\tt SWITCH} allows one to distinguish perfectly between pairs of channels that could not be distinguished perfectly by inserting the corresponding boxes in a circuit in any given order.      
\item  Although the quantum {\tt  SWITCH} can be simulated in an ordinary circuit with only a polynomial slowdown, there is currently no proof that the same can be done for arbitrary maps on product channels.     The general problem of the physical implementation of supermaps on product channels---and, more generally, of higher-order maps---is currently open. 
For this reason,  the assessment of the the computational power of higher-order computation is still open.        
  \end{enumerate} 
 
 The two points above suggests two avenues of future research: 1)  investigating the advantages for information-processing offered by the quantum {\tt  SWITCH} and 2)  investigating the computational power of higher-order computation.  
Based on the analogy with the classical case, it would be natural to expect that all  quantum circuits and higher-order computation are equivalent models, up to a polynomial slowdown. Moreover, if this were not true, the quantum version of the Church-Turing thesis would be disproved, a fact that is deemed to be unlikely by most quantum computer scientists.    However, having a clear-cut proof that higher-order computation is polynomially equivalent to computation in the circuit model is surely desirable, and would probably shed light on the physical realizability of the hierarchy of higher-order transformations.

\section{Conclusions}\label{sec:conclusions}  

Let us start by summarizing the results presented in the paper: We first analyzed the transformations of no-signalling channels that are allowed in quantum mechanics. The transformations considered here take an input no-signalling channel and transform it in a new output channel, respecting convex combinations and positivity and normalization of probabilities. 
First, we showed that transformations of no-signalling channels involving two parties, $A$ and $B$, can be equivalently defined as transformations of product channels $\map A \otimes \map B$, where $\map A$  and $\map B $ are local channels on $A$'s and $B$'s side, respectively.  
Then, we analyzed in detail  a particular example of such a transformation: the {\tt SWITCH} transformation, where an arbitrary pair of channels $(\map A,\map B)$ is transformed in either $\map A \map B$ or in  $\map B \map A$ depending on the state of a control bit.  

The {\tt SWITCH} transformation can be considered as the mathematical description of a quantum computation of higher-order, where the input of the computation is a subroutine  provided as a black box. Such computations are the kind of computations that  would have be included in a complete, quantum version of Church's $\lambda$-calculus (cf. Refs. \cite{vanTonder,selinger,grattage,sv2004,perdrix,arrighi} for an overview of the different extensions of Church's $\lambda$-calculus from the classical to the quantum case).    An important fact of higher-order computations is that, in general, they cannot be implemented by inserting the input black boxes inside an ordinary quantum circuit.   We illustrated this fact in the specific example of the {\tt SWITCH} transformations, showing that no quantum circuit containing a single call to the black boxes  $\map A$ and $\map B$ can implement the transformation {\tt SWITCH} deterministically.   The reason of the impossibility is the fact that the transformation $\tt SWITCH$ is incompatible with any choice of a causal ordering between the boxes $\map A$ and $\map B$. 
In fact, in the paper we showed that realizing the {\tt SWITCH} transformation by simple insertion of the boxes in a given order in a circuit would be equivalent to realizing a time machine, thus violating causality.

Subsequently, discussed four ways around the no-go theorem:  1) allowing access to program states, 2) allowing two queries to the input black boxes, 3)  allowing access to closed timelike curves, and 4)  considering probabilistic simulations.   
Moreover, we discussed a minimal change of the rule for describing the
oracle access to the black boxes   $\map A$ and $\map B$, introducing classical control of causal sequences of
operations, in such a way that the computation of the class of
higher-order functions including the {\tt SWITCH} can be expressed in
circuital terms.  

Finally, we considered the quantum version of the {\tt SWITCH} transformation,  which can be implemented if we allow for  quantum control of causal sequence of operations.   
A complete physical theory of higher-order computation has not been
developed yet, we expect it to reveal unexplored aspects of quantum
theory in a non-fixed causal framework. The quantum switch of boxes is a new primitive that enables computations where the causal structure of the connections can be in a quantum superposition.  
A quantum computational model in which the states of quantum systems can control the
structure of a causal network  suggests a fascinating analogy with a
quantum gravity scenario, in which the space-time geometry can be
entangled with the state of physical systems.

We believe that exhaustive analysis of higher-order transformations in quantum mechanics will
provide some new insight for the formulation of a theory of quantum gravity, within a framework similar to the causaloid framework  of Ref. \cite{causaloid}.  The physical implementation of higher-order functions discussed here has also an
interesting relation to the paradigm of the universe as a quantum
computer \cite{lloyd}.  Indeed, one can wonder what kind of quantum
computer the universe is: It could be a gigantic quantum circuit where information is encoded in the state of many qubits and is processed in time from a spacelike surface to the next, or it could be a
quantum Turing machine, or also be a higher-order computer, that
processes information encoded in transformations (e.g. in scattering
amplitudes) rather than in states.  Even if these three models turn
out to be equivalent from an abstract computational point of view,
they would nevertheless remain very different from the physical one,
as they are based on different physical mechanisms.  Moreover, as we
already mentioned, the third model has still to be completely
formulated: What is presently lacking is a complete physical theory
that characterizes all transformations of boxes that are possible in
nature.  A piece of Quantum Theory has yet to be  explored.

\medskip 

\emph{Acknowledgments.}  We wish to thank the anonymous referee for a
detailed list of comments that helped us to improve the presentation
(in particular, we credit the referee for recommending us to include
in the paper the alternative proof of Theorem \ref{th:noswitch} based
on the quantum comb formalism).  We also thank P. Selinger for
stimulating discussions, during which he independently devised the
realization of the {\tt SWITCH} program by a quantum machine with
movable wires.  G. C. acknowledges support by the National Basic
Research Program of China (973) 2011CBA00300 (2011CBA00302).  Research
at Perimeter Institute for Theoretical Physics is supported in part by
the Government of Canada through NSERC and by the Province of Ontario
through MRI.  Research at U. Penn. has been supported by the
Intelligence Advanced Research Projects Activity (IARPA) via
Department of Interior National Business Center contract number
D11PC20168\cite{notepenn}

\appendix  

\section{Proof of theorem \ref{theo:cp}}  
\label{app:one}
Here we provide the proof details for theorem \ref{theo:cp}.

\noindent {\bf Proof.} Let $\sH_\rC$ be an arbitrary Hilbert space and $ Q\in \Lin (  \sH_{\rA'}  \otimes \sH_{\rA}  \otimes \sH_{\rC}) $   be an arbitrary positive operator.  We want to show that $(\widetilde {\map S}  \otimes \map I_{\rC})  (Q) $ is positive.  

This fact can be proved as follows: Up to a rescaling, $Q$ is the Choi operator of a quantum operation $\map Q\in\QO ( \rA  \to \rA' \rC )$.  Since $\map C_0$ is an internal channel, up to rescaling we also have that 
\begin{align}\label{bound} 
Q  \le C_0  \otimes \rho_0,
\end{align} 
where $\rho_0\in\St (\rC)$ is an arbitrary full-rank state.  
Consider a purification of $C_0  \otimes \rho_0$, given by a Hilbert space $\spc H_\rD$ and a vector $ |V\>  \in \sH_{\rA'} \otimes \sH_{\rA} \otimes \sH_{\rC} \otimes \sH_\rD$ such that 
\begin{align*}
C_0\otimes \rho_0  =   \Tr_{\rD}  [|V\>\< V|].   
\end{align*}
By construction, $ |V\>\< V|$ is the Choi operator of the channel $\map V$ defined as $\map V (\rho): =  \Tr_\rA[   (  I_{\rA'}  \otimes \rho^T  \otimes I_{\rC } \otimes I_\rD)  |V\>\< V|]$ and the channel $\map V$ is an extension of $\map C_0$: 
\begin{align*}
\map C_0 (\rho)  =  \Tr_{\rC \rD}  [ \map V(\rho) ]  \qquad \forall \rho \in \St (\rA).  
\end{align*} 

In other words, defining $\sH_\rE  :  =  \mathbb C$ and $\sH_{\rE'}  :=   \sH_\rC  \otimes \sH_\rD $ as have $\map V \in   \mathsf {Ext}_{\rE \to \rE'}  [  \map C_0 ]$.  Since $\map S$ is a supermap of type $\mathsf S_\rA \to \mathsf S_\rB$ we must have that $(\map S  \otimes \map I_{\rE \to \rE'}) (  \map V) $ is a quantum channel.  
In the Choi representation, this means 
\begin{align}\label{epositivo}
 (  \widetilde{\map S}  \otimes \map I_{\rE'}  \otimes \map I_\rE)  ( |V\>\< V|  )  \ge 0.    
\end{align}
   
Now, since $|V\>$ is a purification of $  C_0 \otimes \rho_0$, Eq. (\ref{bound}) implies there exists a positive operator  $  P  \in   \Lin (\rD)$ such that $Q  =   \Tr_{\rD}   [    (  I_{\rA'\rA \rC}  \otimes P) |V\>\< V| ]$. 
We can then conclude  
\begin{align*}
(\widetilde {\map S}  \otimes \map I_{\rC})  (Q) &  =        (\widetilde {\map S}  \otimes \map I_{\rC})    \left\{  \Tr_{\rD}   [    (  I_{\rA'\rA \rC}  \otimes P) |V\>\< V| ]\right\}  \\
 &  =    \Tr_{\rD}  \left\{       (  I_{\rB'\rB \rC}  \otimes P)       (\widetilde {\map S}  \otimes \map I_{\rC} \otimes \map I_{\rD}  )  [  |V\>\< V| ] \right\}\\
 &  \ge 0,  
   \end{align*} 
the last inequality  following from the relation $   (\widetilde {\map S}  \otimes \map I_{\rC} \otimes \map I_{\rD}  )  [  |V\>\< V| ] \equiv (  \widetilde{\map S}  \otimes \map I_{\rE'}  \otimes \map I_\rE)   [  |V\>\< V| ]   \ge 0 $ [cf. Eq. (\ref{epositivo})].    $\blacksquare$
\section{Alternative proof of the impossibility of a circuit realization of the switch supermap}  

Here we give an alternative proof of Theorem \ref{th:noswitch}, based on the formalism of \emph{quantum combs} \cite{qca,comblong}.   The proof is extremely short once the basic facts about quantum combs are assumed. We include this short proof as an illustration of the power of the quantum comb formalism.    

The formalism of quantum combs consists in a recursive application of the Choi isomorphism.  As  already mentioned, in the Choi representation, any supermap $\map S $  of type $  \Chan  (\rA \to \rA') \to   \Chan  (\rB \to \rB')$, is in 1-to-1 correspondence with a completely positive map $\widetilde {\map S}  :  \Lin  (\sH_{\rA'} \otimes \sH_{\rA }) \to   \Lin  (\sH_{\rB'} \otimes \sH_{\rB })$. 
Applying the Choi isomorphism once more, the completely positive map $\widetilde {\map S}$ is in 1-to-1 correspondence with a positive operator $S  \in  \Lin  (\sH_{\rB'} \otimes \sH_{\rB } \otimes \sH_{\rA'} \otimes \sH_{\rA })$.    
In particular, this construction associates a  supermap $  \map S  $   of type $ \mathsf{Prod}   (\rA\rB \to \rA'\rB') \to \Chan   (\rC \to \rC') $   to a positive operator 
\begin{align*}
S  \in  \Lin  (\sH_{\rC'} \otimes \sH_{\rC } \otimes \sH_{\rA'} \otimes \sH_{\rA }  \otimes \sH_{\rB'} \otimes \sH_{\rB }).\end{align*}

Ref. \cite{comblong} gives necessary and sufficient conditions for the realization of the supermap $\map S$ in a circuit with fixed causal structure:    
precisely, the mapping $\map S:  \map A \otimes \map B  \mapsto   \map S (\map A\otimes \map B)$ can be implemented by a deterministic circuit with $\map A$ preceding $\map B$, namely
\begin{align*}
& \begin{aligned}\Qcircuit @C=1em @R=.7em @! R 
 {&\qw \poloFantasmaCn{\rC}  &\gate{\map S  (  \map A \otimes \map B)}& \qw\poloFantasmaCn{\rC'}  & \qw  }
  \end{aligned}
  =  \\
  &  \\
 & \qquad \qquad   \begin{aligned}\Qcircuit @C=1em @R=.7em @! R 
 {&\qw \poloFantasmaCn{\rC} & \multigate{1}{\map C_1}&\qw \poloFantasmaCn{\rA}  &\gate{\map A}
  &\qw \poloFantasmaCn{\rA'}  &  \multigate{1}{\map C_2}&\qw \poloFantasmaCn{\rB}  &  \gate{\map B}&\qw \poloFantasmaCn{\rB'}  &\multigate{1}{\map C_3}& \qw\poloFantasmaCn{\rC'}  & \qw  \\
     & & \pureghost{\map C_1}&\qw &\qw& \qw& \ghost{\map C_2}&\qw& \qw &\qw &\ghost{\map C_3}&  &\\
      }
  \end{aligned}
\end{align*}
 if and only if there exist positive operators $T \in  \Lin (\sH_\rB\otimes \sH_{\rA'}  \otimes \sH_\rA  \otimes \sH_\rC )$ and  $U  \in  \Lin ( \sH_\rA  \otimes \sH_\rC )$  such that  
 \begin{align}
\nonumber \Tr_{\rC'} [  S]   &   =   I_{\rB'}  \otimes  T\\
\nonumber \Tr_{\rB}  [  T]   &    =  I_{\rA'}  \otimes U  \\
\label{choiab} 
\Tr_{ \rA }  [ U ] &  =  I_\rC.
 \end{align} 
Similarly,  the mapping $\map S:  \map A \otimes \map B  \mapsto   \map S (\map A\otimes \map B)$ can be implemented by a deterministic circuit with $\map B$ preceding $\map A$, namely
\begin{align*}
& \begin{aligned}\Qcircuit @C=1em @R=.7em @! R 
 {&\qw \poloFantasmaCn{\rC}  &\gate{\map S  (  \map A \otimes \map B)}& \qw\poloFantasmaCn{\rC'}  & \qw  }
  \end{aligned}
  =  \\
  &  \\
 & \qquad \qquad   \begin{aligned}\Qcircuit @C=1em @R=.7em @! R 
 {&\qw \poloFantasmaCn{\rC} & \multigate{1}{\widetilde{\map C_1}}&\qw \poloFantasmaCn{\rB}  &\gate{\map B}
  &\qw \poloFantasmaCn{\rB'}  &  \multigate{1}{\widetilde{\map C_2}}&\qw \poloFantasmaCn{\rA}  &  \gate{\map A}&\qw \poloFantasmaCn{\rA'}  &\multigate{1}{\widetilde{\map C_3}}& \qw\poloFantasmaCn{\rC'}  & \qw  \\
     & & \pureghost{\widetilde{\map C_1}}&\qw &\qw& \qw& \ghost{\widetilde{\map C_2}}&\qw& \qw &\qw &\ghost{\widetilde{\map C_3}}&  &\\
      }
  \end{aligned}
\end{align*}
 if and only if there exist positive operators $\widetilde T \in  \Lin (\sH_\rA\otimes \sH_{\rB'}  \otimes \sH_\rB  \otimes \sH_\rC )$ and  $\widetilde U  \in  \Lin ( \sH_\rB  \otimes \sH_\rC )$  such that  
 \begin{align}
\nonumber \Tr_{\rC'} [  S]   &   =   I_{\rA'}  \otimes \widetilde T\\
\nonumber \Tr_{\rA}  [\widetilde  T]   &    =  I_{\rB'}  \otimes \widetilde U  \\
\label{choiba} 
\Tr_{ \rB }  [ \widetilde U ] &  =  I_\rC.
 \end{align} 
 
Once these facts are known, the proof becomes  very quick:  

{\bf Proof of theorem \ref{th:noswitch}.}     
Denoting by $E$ the rank-one operator $E  :=   |I\>\<  I  |$, where $ |I\>  : =  \sum_{n}  |n\>|n\>$, and suitably reordering the Hilbert spaces, the switch supermap $\map S$ has Choi operator 
\begin{align*}
S    =&   {P_0}_{\rQ}   \otimes  Z_0  + {P_1}_{\rQ}     \otimes  Z_1 
 \end{align*}    
with $Z_0$ and $Z_1$ being the Choi operators of the supermaps $\map Z_0$ and $\map Z_1$ defined in Eqs. (\ref{zero}) and (\ref{uno}) 
\begin{align*}
Z_0  :  =    E_{ \rC'   \rB' }  \otimes E_{\rB \rA'}  \otimes E_{\rC\rA}  \\
Z_1  :  =   E_{ \rC'   \rA' }  \otimes E_{\rA \rB'}  \otimes E_{\rC\rB} .
\end{align*}
Now, $Z_0$ satisfies the condition (\ref{choiab}) and $Z_1$ satisfies the condition (\ref{choiba}), but their sum $S  =  {P_0}_{\rQ}   \otimes  Z_0  + {P_1}_{\rQ}     \otimes  Z_1 $ does not satisfy any of these conditions.  Hence, the supermap $\map S$   cannot be realized by inserting $\map A$ and $\map B$ in a quantum circuit in a definite order. $\blacksquare$

\end{document}

%% file: myQcircuit.tex
\usepackage[matrix,frame,arrow]{xy}
\usepackage{amsmath}

\newcommand{\ket}[1]{\left\vert{#1}\right\rangle}
\newcommand{\qw}[1][-1]{\ar @{-} [0,#1]}
\newcommand{\qwx}[1][-1]{\ar @{-} [#1,0]}
\newcommand{\gate}[1]{*{\xy *+<.6em>{#1};p\save+LU;+RU **\dir{-}\restore\save+RU;+RD **\dir{-}\restore\save+RD;+LD **\dir{-}\restore\POS+LD;+LU **\dir{-}\endxy} \qw}
\newcommand{\measureD}[1]{*{\xy*+=+<.5em>{\vphantom{\rule{0em}{.1em}#1}}*\cir{r_l};p\save*!R{#1} \restore\save+UC;+UC-<.5em,0em>*!R{\hphantom{#1}}+L **\dir{-} \restore\save+DC;+DC-<.5em,0em>*!R{\hphantom{#1}}+L **\dir{-} \restore\POS+UC-<.5em,0em>*!R{\hphantom{#1}}+L;+DC-<.5em,0em>*!R{\hphantom{#1}}+L **\dir{-} \endxy} \qw}
\newcommand{\multimeasureD}[2]{*+<1em,.9em>{\hphantom{#2}}\save[0,0].[#1,0];p\save !C *{#2},p+LU+<0em,0em>;+RU+<-.8em,0em> **\dir{-}\restore\save +LD;+LU **\dir{-}\restore\save +LD;+RD-<.8em,0em> **\dir{-} \restore\save +RD+<0em,.8em>;+RU-<0em,.8em> **\dir{-} \restore \POS !UR*!UR{\cir<.9em>{r_d}};!DR*!DR{\cir<.9em>{d_l}}\restore \qw}
\newcommand{\control}{*!<0em,.025em>-=-{\bullet}}
\newcommand{\ctrl}[1]{\control \qwx[#1] \qw}
\newcommand{\multigate}[2]{*+<1em,.9em>{\hphantom{#2}} \qw \POS[0,0].[#1,0];p !C *{#2},p \save+LU;+RU **\dir{-}\restore\save+RU;+RD **\dir{-}\restore\save+RD;+LD **\dir{-}\restore\save+LD;+LU **\dir{-}\restore}
\newcommand{\ghost}[1]{*+<1em,.9em>{\hphantom{#1}} \qw}
\newcommand{\ustick}[1]{*!D!<0em,-.5em>=<0em>{#1}}
\newcommand{\Qcircuit}[1][0em]{\xymatrix @*=<#1>} % @*[o]
\newcommand{\pureghost}[1]{*+<1em,.9em>{\hphantom{#1}}}
\newcommand{\multiprepareC}[2]{*+<1em,.9em>{\hphantom{#2}}\save[0,0].[#1,0];p\save !C
  *{#2},p+RU+<0em,0em>;+LU+<+.8em,0em> **\dir{-}\restore\save +RD;+RU **\dir{-}\restore\save
  +RD;+LD+<.8em,0em> **\dir{-} \restore\save +LD+<0em,.8em>;+LU-<0em,.8em> **\dir{-} \restore \POS
  !UL*!UL{\cir<.9em>{u_r}};!DL*!DL{\cir<.9em>{l_u}}\restore}
\newcommand{\poloFantasmaCn}[1]{{{}^{#1}_{\phantom{#1}}}}

%% file: beyondqc_stop.bbl
\begin{thebibliography}{99}
\bibitem{disclaimer}   As we will mention later in the paper, higher-order quantum computation does offer some advantages in information processing tasks such as the discrimination of no-signalling channels. These advantages, however, do not imply that the higher-order quantum model would change complexity classes  with respect to the quantum circuit model.
\bibitem{notepenn} The U.S. Government is authorized to reproduce and
  distribute reprints for Governmental purposes notwithstanding any
  copyright annotation thereon. Disclaimer: The views and conclusions
  contained herein are those of the authors and should not be
  interpreted as necessarily representing the official policies or
  endorsements, either expressed or implied, of IARPA, DoI/NBC, or the
  U.S. Government.
\end{thebibliography}
